\documentclass{jfm}

\usepackage{amssymb}
\usepackage{amsbsy}
\usepackage{mathrsfs,mathcomp}
\usepackage{amsmath}
\usepackage{amsfonts}
\usepackage{graphicx}
\usepackage{upgreek}
\usepackage{natbib}
\usepackage[english]{babel}
\usepackage{ifthen}
\usepackage{booktabs}
\usepackage{jabbrv}

\newcommand{\comb}[1]{#1} 				

\newcommand{\degree}{\ensuremath{^\circ}} 				
\newcommand{\f}[1]{\ensuremath{\mathrm{f}_{#1}}} 		
\newcommand{\Ro}{\ensuremath{R_0}} 						
\newcommand{\Rmax}{\ensuremath{R_\mathit{max}}} 		
\newcommand{\tmax}{\ensuremath{t_\mathit{max}}} 		
\newcommand{\tauc}{\ensuremath{\tau_c}} 				
\newcommand{\taup}{\ensuremath{\tau_p}} 				
\newcommand{\taui}{\ensuremath{\tau_i}} 				
\newcommand{\taue}{\ensuremath{\tau_e}} 				
\newcommand{\pe}{\ensuremath{p_e}} 						
\newcommand{\tnorm}{\ensuremath{t/\tau_c}} 				
\newcommand{\taub}{\ensuremath{t_s}} 					
\newcommand{\hattaub}{\ensuremath{\hat{t}_s}} 			
\newcommand{\taul}{\ensuremath{t_r}} 					
\newcommand{\kl}{\ensuremath{k_r}} 						
\newcommand{\kb}{\ensuremath{k_s}} 						
\newcommand{\hatkb}{\ensuremath{\hat{k}_s}} 			
\newcommand{\We}{\ensuremath{\mathit{We}}}				
\newcommand{\Reyn}{\ensuremath{\mathit{Re}}}				
\newcommand{\Oh}{\ensuremath{\mathit{Oh}}}				
\newcommand{\Wed}{\ensuremath{\mathit{We}_d}}			
\newcommand{\Ecm}{\ensuremath{E_\mathit{k,cm}}}			
\newcommand{\Ed}{\ensuremath{E_\mathit{k,d}}} 			
\newcommand{\EToE}{\ensuremath{\Ed/\Ecm}}				
\newcommand{\EToEFrac}{\ensuremath{\frac{\Ed}{\Ecm}}}	
\newcommand{\ELaser}{\ensuremath{E_L}}			
\newcommand{\lLaser}{\ensuremath{\lambda_L}}	
\newcommand{\Eabs}{\ensuremath{E_\mathit{abs}}}			
\newcommand{\Eth}{\ensuremath{E_\mathit{th}}}			
\newcommand{\Hdelta}{\ensuremath{\Delta H}}				
\newcommand{\Nrl}{\ensuremath{N_r}}						
\newcommand{\Nsl}{\ensuremath{N_s}}						
\newcommand{\hb}{\ensuremath{h_s}}						
\newcommand{\Fig}[1][]{%
\ifthenelse{\equal{#1}{}}{figure~}{figure#1~}%
}
\newcommand{\Chap}[1][]{%
\ifthenelse{\equal{#1}{}}{\S\,}{\S#1\,}%
}
\newcommand{\Eq}[1][]{%
\ifthenelse{\equal{#1}{}}{}{#1~}%
}
\newcommand{\Tab}[1][]{%
\ifthenelse{\equal{#1}{}}{table~}{table#1~}%
}

\makeatletter
\def\bbl@set@language#1{%
  \edef\languagename{%
    \ifnum\escapechar=\expandafter`\string#1\@empty
    \else\string#1\@empty\fi}%
  \@ifundefined{babel@language@alias@\languagename}{}{%
    \edef\languagename{\@nameuse{babel@language@alias@\languagename}}%
  }%
  \select@language{\languagename}%
  \expandafter\ifx\csname date\languagename\endcsname\relax\else
    \if@filesw
      \protected@write\@auxout{}{\string\select@language{\languagename}}%
      \bbl@for\bbl@tempa\BabelContentsFiles{%
        \addtocontents{\bbl@tempa}{\xstring\select@language{\languagename}}}%
      \bbl@usehooks{write}{}%
    \fi
  \fi}
\newcommand{\DeclareLanguageAlias}[2]{%
  \global\@namedef{babel@language@alias@#1}{#2}%
}
\makeatother

\DeclareLanguageAlias{en}{english}

\makeatletter
\let\Hy@backout\@gobble
\makeatother

\newcommand{\refl}[1]{\protect\includegraphics{#1}}
\usepackage{xcolor}
\definecolor{mycolor1}{rgb}{0.00000,0.44700,0.74100}%
\definecolor{mycolor2}{rgb}{0.85000,0.32500,0.09800}%
\definecolor{mycolor3}{rgb}{0.92900,0.69400,0.12500}%
\definecolor{mycolor4}{rgb}{0.49400,0.18400,0.55600}%
\definecolor{mycolor5}{rgb}{0.46600,0.67400,0.18800}%
\definecolor{mycolor6}{rgb}{0.30100,0.74500,0.93300}%
\definecolor{mycolor7}{rgb}{0.63500,0.07800,0.18400}%

\title[Drop fragmentation by laser-pulse impact]{Drop fragmentation by laser-pulse impact}
\author[Klein et al.]{Alexander L. Klein$^1$, Dmitry Kurilovich$^{2,3}$, Henri Lhuissier$^4$, Oscar O. Versolato$^2$, Detlef Lohse$^1$, Emmanuel Villermaux$^{5,6}$ and Hanneke Gelderblom$^{1,7}$\footnote{\label{cor}\email{h.gelderblom@tue.nl}}}
 
\affiliation{
$^1$Physics of Fluids Group, Max Planck Center Twente for Complex Fluid Dynamics, J.M.\ Burgers Center, and MESA+ Center for Nanotechnology, Department of Science and Technology, University of Twente, P.O.\ Box 217, 7500 AE Enschede, The Netherlands.\\[\affilskip]
$^2$Advanced Research Center for Nanolithography (ARCNL), \\Science Park 106, 1098 XG Amsterdam, The Netherlands.\\[\affilskip]
$^3$Department of Physics and Astronomy, and LaserLaB,\\ Vrije Universiteit, De Boelelaan 1081, 1081 HV Amsterdam, The Netherlands.\\[\affilskip]
$^4$Aix Marseille Universit\'e, CNRS, IUSTI, Marseille, France.\\[\affilskip]
$^5$ Aix Marseille Universit\'e, CNRS, Centrale Marseille, IRPHE, Marseille, France.\\[\affilskip]
$^6$Institut Universitaire de France, Paris, France.\\[\affilskip]
$^7$Department of Applied Physics,
Eindhoven University of Technology,\\ Den Dolech 2, 5600 MB, Eindhoven, Netherlands.
}
\pubyear{??}
\volume{??}
\pagerange{??--??}
\date{?? and in revised form ??}

\begin{document}
\maketitle

\begin{abstract}
We study the fragmentation of a liquid drop that is hit by a laser pulse. The drop expands into a thin sheet that breaks by the radial expulsion of ligaments from its rim and the nucleation and growth of holes on the sheet. By combining experimental data from two liquid systems with vastly different time- and length scales we show how the early-time laser-matter interaction affects the late-time fragmentation. We identify two Rayleigh--Taylor instabilities of different origins as the prime cause of the fragmentation and derive scaling laws for the characteristic breakup time and wavenumber. The final web of ligaments results from a subtle interplay between these instabilities and deterministic modulations of the local sheet thickness, which originate from the drop deformation dynamics and spatial variations in the laser-beam profile.
\end{abstract}

\section{Introduction}\label{frag:Introduction}
\begin{figure}
  \centering
  \includegraphics{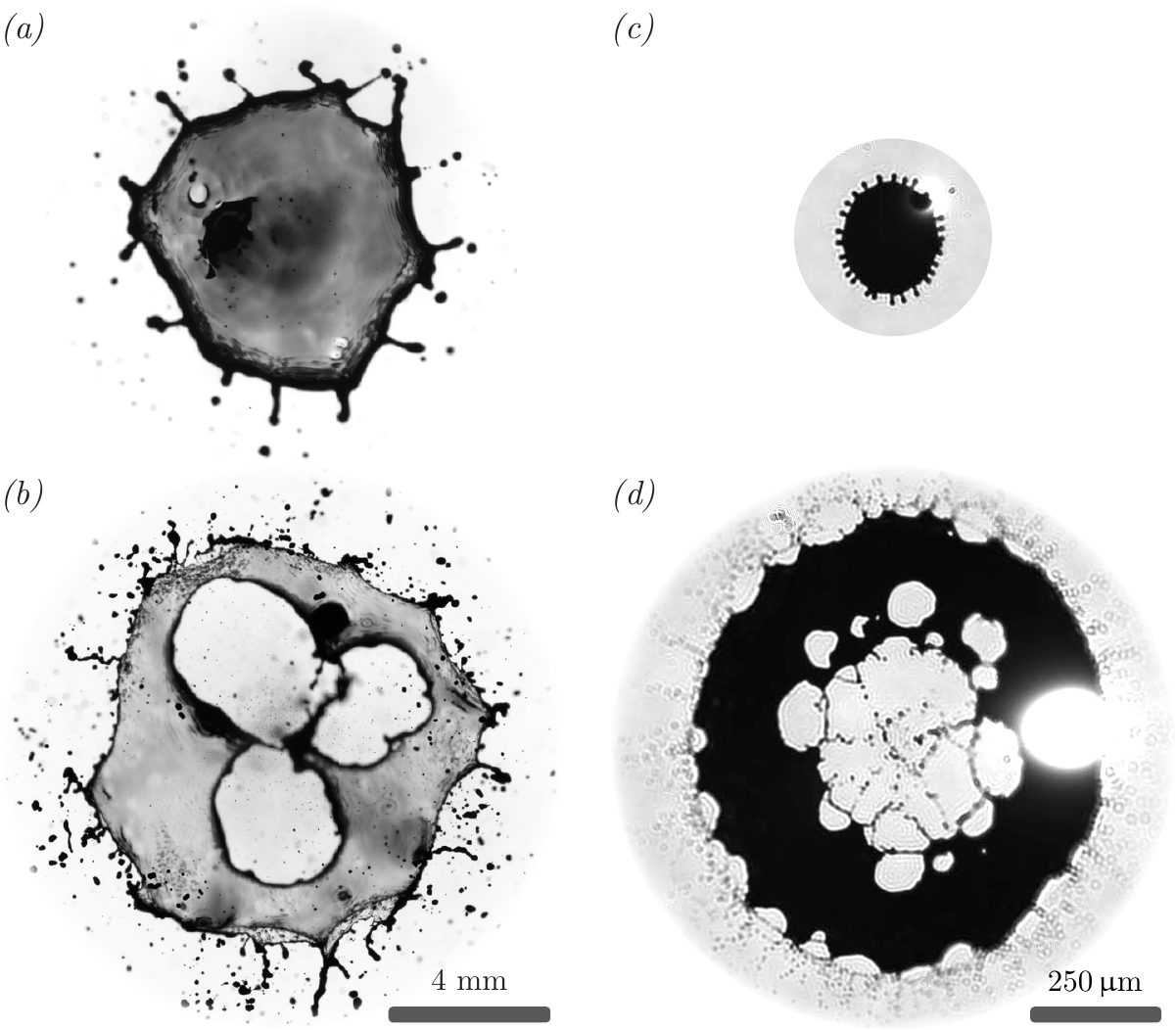}
  \caption{Fragmentation of drops of methyl ethyl ketone (MEK, \textit{a},~\textit{b}) and liquid tin (\textit{c},~\textit{d}) following the impact of a laser pulse. The drops are accelerated by the laser impact and deform into thin liquid sheets that break by the radial expulsion of ligaments (\textit{a},~\textit{c}) and by the nucleation and growth of holes (\textit{b},~\textit{d}). The two drops differ in length scale and in propulsion mechanism. The millimetre-sized MEK drop is accelerated by the local boiling of MEK and the micron-sized tin drop by an expanding and glowing plasma cloud, which is visible as a white spot in (\textit{c},~\textit{d}).\label{fig:Introduction}}
\end{figure}

	The impact of a nanosecond laser-pulse onto a opaque liquid drop induces large-scale deformation and eventually fragmentation of the liquid. Figure~\ref{fig:Introduction} shows how the laser impact causes a spherical drop to deform into a thin liquid sheet that later on breaks into a set of ligaments and smaller drops. Our previous work \citep{gelderblom_drop_2016} has addressed the drop deformation in this early phase in detail. The subsequent laser-induced fragmentation is the subject of the present study. Understanding this fragmentation is of key importance for the development of laser-produced plasma light sources for extreme ultraviolet (EUV) nanolithography, in which a dual laser-pulse impact on a tin drop triggers the emission of EUV light by ionising the tin~\citep{banine_physical_2011}. A first pulse shapes the drop into a thin sheet that is ionised by the second, high-energy pulse.
The dispersion and exposure of the liquid tin to the second pulse, which is crucial for the efficient generation of EUV light, is directly determined by the mechanics of deformation and fragmentation of the sheet.
	
The fragmentation of a drop has been studied extensively for mechanical impacts onto a solid substrate or a pillar \citep[see e.g.][]{roisman_horvat_2006,xu_splashing_2007,villermaux_drop_2011, riboux_diameters_2015,bourouiba_2018}. For these impacts the breakup results from the Rayleigh-Taylor and Rayleigh-Plateau instabilities of the rim bordering the radially expanding drop. For a laser pulse impacting a transparent liquid the fragmentation has been shown to result from explosive vaporisation \citep{kafalas_fog_1973}, plasma bubble formation \citep{lindinger_time-resolved_2004}, the generation of shock waves \citep{stan_2016}, rapid expansion of an enclosed explosive gas  \citep{vledouts_explosive_2016}, or acoustic cavitation \citep{gonzalez_avila_fragmentation_2016}. By contrast, when a laser pulse impacts an opaque liquid drop, the laser-liquid interaction remains restricted to a superficial layer. The local energy deposition induces a phase change that gives rise to a strong recoil pressure on the surface of the drop. For ultrashort (i.e.~femto- and picosecond) laser pulses this violent recoil pressure induces shock waves, cavitation and explosive fragmentation of the drop \citep{grigoryev_2018, kurilovich_2018}. In the present study, we consider the more moderate regime of nanosecond laser pulses. In this case the response of the drop occurs on a timescale much larger than the acoustic time and can be considered incompressible \citep{reijers_pulses_2017}. 
As a result of the recoil pressure the drop is propelled forward, deforms, and eventually fragments \citep{klein_drop_2015}. The laser-induced drop deformation primarily depends on the Weber number \citep{gelderblom_drop_2016}
\begin{equation}
	\We = \frac{\rho \Ro U^2}{\gamma},\label{eq:Weber}
\end{equation}
where $\rho$ is the liquid density, \Ro\ the initial drop radius, $\gamma$ the surface tension, and $U$ the centre-of-mass velocity of the drop, which is determined by the laser-pulse energy \citep{klein_drop_2015}. As we will show, this Weber number is also the key parameter governing fragmentation of the drop.

	We study this laser-induced fragmentation experimentally using two liquids: a dyed solvent and liquid tin. The former has many practical experimental advantages that will be discussed below, whereas the latter is inspired by the EUV lithography application. The combination of the two systems allow us to explore both a broad range of \We\ and the effect of the differences in the laser-matter interaction. The dyed solvent drops are propelled by a local boiling and vapour expulsion \citep{klein_drop_2015}, whereas the tin drops are pushed by an expanding plasma cloud \citep{kurilovich_plasma_2016}.

	In both systems two types of breakup contribute to the fragmentation as shown in \Fig\ref{fig:Introduction}: the radial expulsion of ligaments from the rim of the sheet formed by the flattened drop (\Fig\ref{fig:Introduction} \textit{a}, \textit{c}) and the nucleation of holes on the thin sheet itself (\Fig\ref{fig:Introduction} \textit{b}, \textit{d}). These phenomena have been observed in other experimental systems, e.g.~after the impact of a drop onto a solid obstacle \citep{villermaux_drop_2011} or after the impact of a shock wave onto a thin liquid film \citep{bremond_bursting_2005}. 
The present situation deviates from these studies in two important aspects. First, the laser impact allows to separate the timescales of the drop acceleration and of the subsequent deformation and fragmentation \citep{gelderblom_drop_2016}, which are naturally coupled for the impact on a solid. Second, hole nucleation takes place on an expanding liquid sheet that is formed by the impact of a laser pulse with a certain beam profile, whereas the fixated soap film used by \citet{bremond_bursting_2005} is of constant thickness and hit by a uniform shock front. These differences turn out to have important consequences for the fragmentation dynamics.

	The details of the liquid systems and experimental setups are described in \S\ref{frag:Setup}. In \S\ref{frag:Observation} we qualitatively discuss the experimental observations and illustrate the different breakup phenomena. The deformation of the drop into a sheet is summarised in \S\ref{frag:Expansion} and compared to an existing model.
With a description of the drop kinematics at hand, we analyse the breakup of the sheet rim in~\S\ref{frag:RimBreakup} and the hole nucleation in the sheet in~\Chap\ref{frag:SheetBreakup}. In \S \ref{fragments} the resulting fragment size distributions are discussed qualitatively and a phase diagram outlining the different fragmentation regimes is presented.

\section{Experimental setups}\label{frag:Setup}
	
	We perform experiments with two liquid systems having vastly different length scales. The first system consists of $0.9$-mm methyl-ethyl-ketone drops dyed with Oil-Red-O, which we from now on refer to as MEK drops. A detailed characterisation of the MEK solutions is given in \citet{klein_apparatus_2017}.
The second system consist of $24$-$\umu$m tin drops. We either use pure liquid tin ($99.995\,\%$ purity by Goodfellow), which is motivated by the industrial application in EUV light sources, or an eutectic indium-tin alloy (50In--50Sn, $99.9\,\%$ purity by Indium Corporation) with a conveniently low melting point. Since both the pure tin and the indium-tin alloy are almost equivalent in terms of atomic mass, density and surface tension, we use them interchangeably in this work and refer to them as the tin system, in contrast to the MEK system.
\begin{table}
	\centering
	\begin{tabular}{llcc} %
		& \emph{Description} & MEK & Tin \\[3pt] %
		$T$ 		& liquid temperature ($\degree\mathrm{C}$) & 20 & 260 \\
		$\rho$ 		& liquid density ($\mathrm{kg}\,\mathrm{m}^{-3}$) & 805 & 6968 \\
		$\nu$     	& liquid viscosity ($\mathrm{m}^2\,\mathrm{s}^{-1}$) & $0.53 \times 10^{-6}$ & $0.27 \times 10^{-6}$\\
		$\gamma$ 	& surface tension ($\mathrm{N}\,\mathrm{m}^{-1}$) & 0.025 & 0.544 \\[3pt]
		\Ro 		& initial drop radius ($\mathrm{m}$) & $0.9\times 10^{-3}$ & $24\times 10^{-6}$ \\
		\tauc 		& capillary timescale ($\mathrm{s}$) & $5 \times 10^{-3}$ & $13 \times 10^{-6}$\\
		\taui 		& inertial timescale ($\mathrm{s}$) & $\sim 10^{-4}$ & $\sim 10^{-6}$\\
		\taue 		& propulsion timescale ($\mathrm{s}$) & $\sim 10^{-5}$ & $\sim 10^{-8}$\\
		\taup 		& laser duration (FWHM) ($\mathrm{s}$) & $5 \times 10^{-9}$ & $10 \times 10^{-9}$ \\
		\lLaser 	& laser wavelength ($\mathrm{nm}$) & 532 & 1064 \\
		--- & propulsion mechanism & vapour-driven & plasma-driven\\[3pt]%
		\We 		& Weber number range & 90--2000 & 5--18500\\
		\Reyn 		& Reynolds number range  & 3000--14\,000 & 400--22000\\
		\Oh 		& Ohnesorge number  & $\ll 1$ & $\ll 1$ %
	\end{tabular}
		\caption{\label{tab:Systems}Characteristics of the two experimental systems. The MEK system uses a drop of a solution of dye Oil-Red-O in methyl ethyl ketone and a nitrogen environment at ambient temperature (for details about the dye manufacturer see \citet{klein_apparatus_2017}). The second system consists of liquid tin at an elevated temperature in a vacuum environment (manufacturer of the liquids given in the text). The laser-pulse duration \taup\ is quantified in both systems by the full width at half maximum (FWHM).}
\end{table}

	Table \ref{tab:Systems} gives an overview of the characteristic parameters of the two systems. In both systems, the laser-pulse duration $\taup$ and timescale for the ejection of matter $\taue$ are strongly decoupled from the timescales of the subsequent fluid dynamic response \citep{klein_drop_2015}, i.e., the inertial time $\taui\sim \Ro/U$, on which the drop propels and deforms, and the capillary time~$\tauc = {(\rho \Ro^3/\gamma)}^{1/2}$, on which the deformation is slowed down by surface tension, according to 
\begin{equation}
	\tau_p,\, \taue \ll \taui < \tauc.\label{eq:TimeScales_Frag}
\end{equation}
As a consequence, the two systems show a similar fluid dynamic response despite the differences in early-time laser-matter interaction. Also, for both system the viscous effects are negligible since the Ohnesorge number $\Oh = \nu\sqrt{\rho/\gamma R_0}\ll 1$. Hence, the Weber number is the key dimensionless number that governs the fluid dynamic response of the drop.

	MEK and tin drops are studied in two different setups providing the same impact configuration as detailled in \S \ref{sec:experiment}. Each system offers respective advantages for our analysis. On the one hand, the millimetre-sized MEK drops expand into semi-transparent sheets that are accessible by high-resolution visualisation. In addition, the relatively long deformation timescale of the sheets $\tau_c$ (see Table \ref{tab:Systems}) allows for high-speed recordings of individual breakup events, which is crucial for the analysis given their stochastic nature \citep{villermaux_fragmentation_2007}. On the other hand, micrometre-sized tin drops achieve much higher Weber numbers under highly symmetric impact conditions that are free of azimuthal modulations in the propulsion mechanism, as will be explained \S \ref{frag:BeamProfile}.

\subsection{Key concept of the experiment}\label{sec:experiment}

\begin{figure}
	\centering 
	\includegraphics{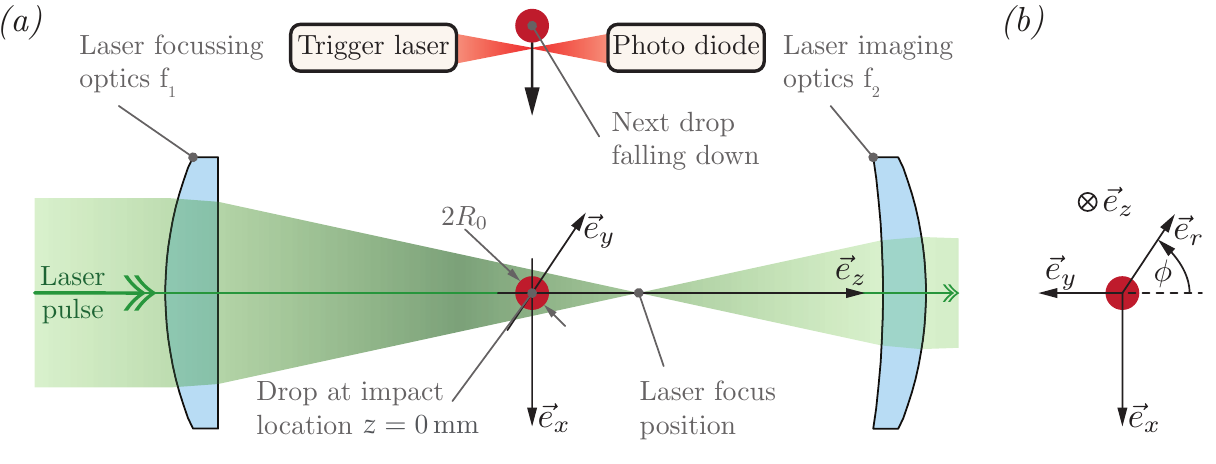}
	\caption{(\textit{a}) Side-view sketch of the drop-impact experiment at the moment of laser impact ($t=0$). The laser pulse is focused with a lens of effective focal length~\f{1}, hits the drop, and is redirected with an imaging lens \f{2} onto a charge-coupled device (CCD) for its characterization. The drop center at the impact location defines the origin of our coordinate system, which is sketched in (\textit{b}) from a back-view ($\vec{e}_z$-direction). The experiment is repeated each time a new drop reaches $x=0$.\label{fig:frag:Chamber}}
\end{figure}

	In both setups, a drop falls down to the laser-impact position while it relaxes to a spherical shape with radius \Ro\ (see \Fig\ref{fig:frag:Chamber}). On its route the drop intercepts a horizontal light sheet that generates a synchronization signal. This signal is used to trigger the impact of the drop by the main laser, the acquisition of the laser pulse energy $\ELaser$ by an energy meter, as well as a beam profiler and two cameras for the visualisation. The complete arrangement of the synchronization laser, photodiode and equipment for the drop generation can be moved in the $yz$-plane to adjust the drop trajectory relative to the laser focus. The delay between the trigger and the laser pulse is tuned to align the drop with the pulse. The pulse  enters from the left through a focusing lens \f{1}, hits the drop at $x=y=z=0$ and exits to the right through the imaging lens \f{2}, which allows to characterise the pulse and the drop irradiation (see \S\ref{frag:BeamProfile}).

	The response of the drop to the laser impact is observed from two orthogonal views: the side-view, aligned with $\vec{e}_y$, and the back-view, aligned with the pulse and drop propagation ($\vec{e}_z$), see \Fig\ref{fig:frag:Chamber}\,(\textit{b}). Stroboscopic image sequences are obtained by performing a new impact experiment and incrementing the time delay between the laser impact and the pulsed light source that illuminates the scene for each image. Image analysis yields the drop centre-of-mass position in all three coordinate directions as a function of time, which is used to calculate the velocity $U$ along $\vec{e}_z$. For $t>\tau_e$ this velocity is constant \citep{klein_drop_2015}. The equivalent sheet radius $R$ is determined as the radius of the circle with the same projected area as the sheet (in the $xy$-plane). Experiments that suffer considerably from a laser-to-drop misalignment or variations in the laser energy are excluded of our analysis. We typically filter out the worst $10\,\%$ of all experimental realisations. 

	The technical equipments used for the MEK and tin experiments differ and are described in detail in \citet{klein_apparatus_2017} and \cite{kurilovich_plasma_2016}, respectively. In the current work, the backlighting in the tin setup has been improved: a pulsed dye-laser pumped by the second harmonic wavelength of a Nd:YAG laser emitting an approximately 5\,ns pulse of 560\,nm light with a spectral width of $\sim4\,\mathrm{nm}$ is used. This lighting reduces the detrimental effects due  to temporal coherence, such as speckle, which enables the visualisation of small features of the expanding tin sheets.

\subsection{Laser-matter interaction}\label{frag:BeamProfile}
	
	The nature of the the laser-matter interaction is a key difference between the two systems. As this interaction will turn out to be important for understanding the late-time fragmentation of the sheet (see \S\ref{frag:SheetBreakup}), we summarise the difference here, while more details can be found in \cite{klein_drop_2015,klein_apparatus_2017} and \cite{kurilovich_plasma_2016}.

	In the MEK system the driving mechanism for the drop acceleration and deformation is a local boiling that is induced by the absorption of laser energy in a superficial layer of the drop. The thickness $\delta$ of this layer is determined by the amount of dye dissolved in the liquid and the absorption coefficient of the dye at the laser wavelength \citep{klein_apparatus_2017}. The laser-dye combination is chosen such that $\delta/\Ro \sim 10^{-2} \ll 1$, which is also the case for the opaque tin drops \citep{cisneros_dielectric_1982-1}. On a timescale~$\taue\sim 10\,\upmu\mathrm{s}$ this layer vaporises and is ejected at the thermal velocity $u$. On the same timescale, the resulting recoil pressure~$\pe$ accelerates the remainder of the drop to the centre-of-mass velocity \citep{klein_drop_2015}
\begin{equation}
	U\sim \frac{\Eabs-\Eth}{\rho\, \Ro^3\, \Hdelta}\,u,\label{eq:PropulsionSpeed}
\end{equation}
where $\Eabs$ is the energy absorbed by the drop, $\Eth$ is the threshold energy that is needed to heat the liquid layer to the boiling point, and $\Hdelta$ is the latent heat of vaporisation. The scaling law~(\ref{eq:PropulsionSpeed}) motivates our choice to use the solvent methyl ethyl ketone (MEK) for the current study. The low value of $\Hdelta$ results in large drop velocities for a given laser energy, which translates into a large range of accessible Weber numbers. 

	For the tin drops the local fluence of the laser exceeds the ionisation threshold. A plasma forms within a fraction of the laser-pulse duration $\taup=10\,\mathrm{ns}$, after which inverse-brems\-strah\-lung absorption strongly decreases the initially high reflectivity of the metallic surface to negligible values \citep{kurilovich_plasma_2016}. Any further laser radiation is absorbed by the plasma cloud. The expanding plasma exerts a pressure~$\pe$ on the drop surface that accelerates the drop. The timescale of this acceleration is set by the plasma dynamics, which is of the same order as the laser-pulse duration, i.e., $\taue \sim \taup = 10\,\mathrm{ns}$. Hence, as for the vapour-driven MEK drops, the tin drops are propelled by a short recoil pressure~$\pe$. Similarly, the centre-of-mass velocity~$U$ for tin scales with the absorbed energy, that is $U \sim {(\Eabs-\Eth)}^{0.59}$, where \Eabs, \Eth, and the exponent now have their origin in the plasma dynamics \citep{kurilovich_plasma_2016}. 

\begin{figure}
	\centering
	\includegraphics{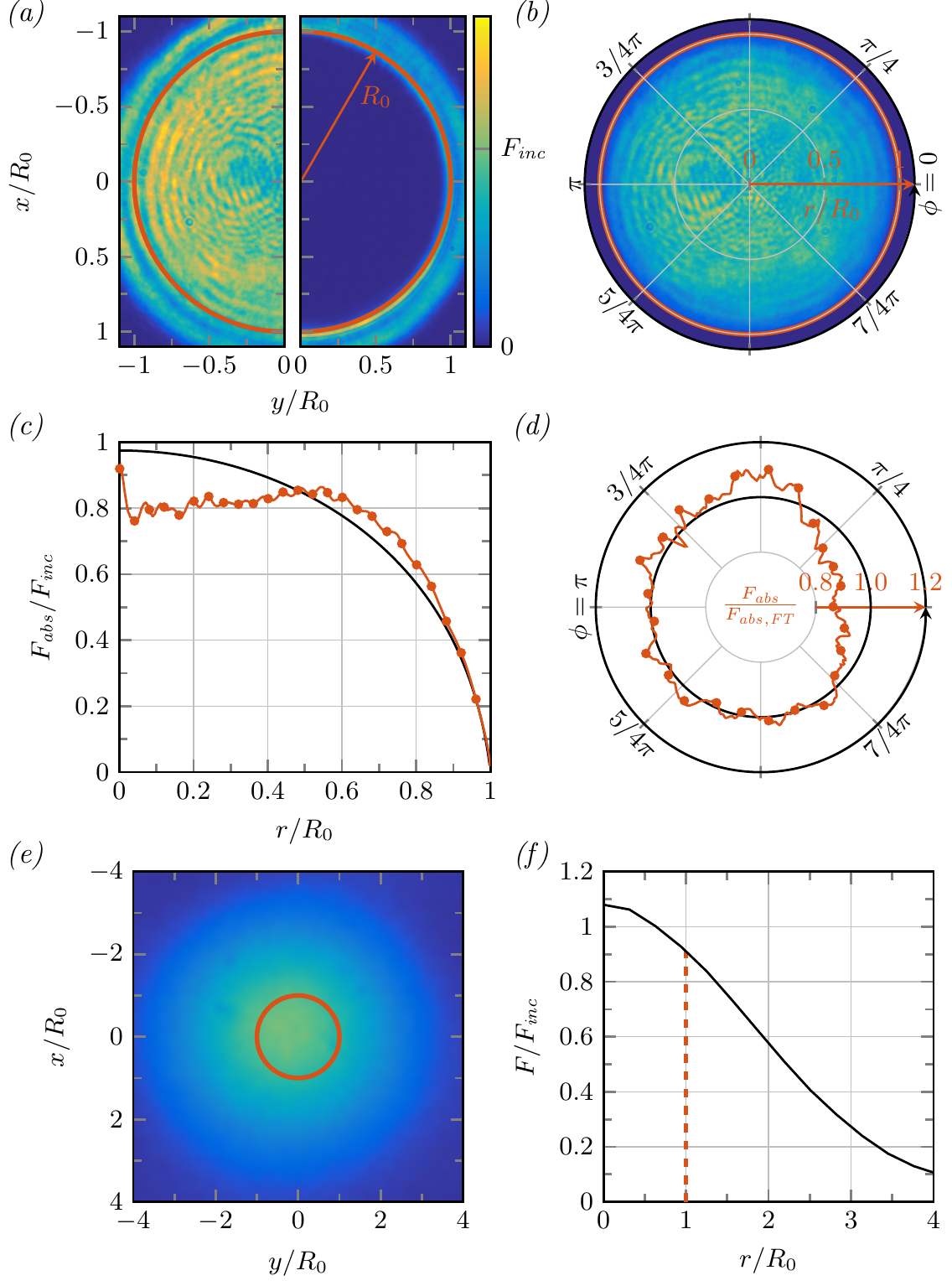} 
	\caption{(\textit{a}) Planar laser-beam profile for the MEK system as recorded without a drop ($y/\Ro \le 0$) and with a drop (for $y/\Ro \ge 0$). The latter yields the drop radius $\Ro$ and position in the beam profile as indicated by the red solid line. The quantity $F_\mathit{inc}$ is the average fluence incident on the drop as given by (\ref{eq:frag:IntegrateBeam}).
	(\textit{b}) Fluence $F_\mathit{abs}$ absorbed by the drop considering the losses due to Fresnel reflection at the liquid-air interface~\citep{hecht_optics_2002}. (\textit{c-d}) Laser profile (red solid line) in radial (\textit{c}, azimuthally averaged) and azimuthal directions (\textit{d}, radially averaged) obtained from $\sim100$ recordings of the planar profile. 
	The black solid line indicates a perfect flat-top beam profile (denoted as $F_\mathit{abs,FT}$ in \textit{d}). (\textit{e}) Planar laser-beam profile measured for the tin system. The red solid line indicates the drop location on impact. The colorbar is the same as in (\textit{a}), which illustrates the smoother and more uniform irradiation of the drop compared to the MEK case. (\textit{f}) Radial beam profile obtained from (\textit{e}).\label{fig:frag:Beamprofile}}
\end{figure}

	To obtain the local laser fluence experienced by the drops, we characterise the laser beam in each system in absence of the drop using the lens \f{2} that images the incident fluence $F_\mathit{inc}$ in the impact plane (\Fig\ref{fig:frag:Chamber}). First, the total radiative energy $\ELaser$ of the pulse is measured with an energy meter capturing the whole beam of light.
Second, a CCD records the relative fluence $f(x,y,z=0)$, which is translated into absolute terms using
\begin{eqnarray}
	F_\mathit{inc} = F(x,y,z=0) &=& \frac{f(x,y,z=0)}{\int f{(x,y,z=0)} \mathrm{d}x\mathrm{d}y} \ELaser.\label{eq:frag:IntegrateBeam}
\end{eqnarray}
Using the position of the drop on impact obtained with the same CCD, we then compute the fluence~$F_\mathit{abs}$ that is actually absorbed by the drop as shown in \Fig\ref{fig:frag:Beamprofile}\,(\textit{b}). From the same arguments underlying (\ref{eq:PropulsionSpeed}), the local recoil pressure~$\pe$ on the drop surface is expected to follow the spatial variations in $F_\mathit{abs}$ according to
\begin{equation}
\pe(r,\phi) \sim \frac{F_\mathit{abs}(r,\phi) - F_\mathit{th}}{\Hdelta}\, \frac{u}{\taue}.\label{eq:PropulsionPressure}
\end{equation} 

	Given the spatial variation in fluence observed in \Fig\ref{fig:frag:Beamprofile}\,(\textit{d}) this suggests that the MEK drops are subject to a driving force that varies along the azimuthal direction~$\phi$ by about $\pm 10\%$. Importantly, since $f$ is found to be independent of $\ELaser$, these spatial variations in the driving force are independent of \ELaser\ and fixed in the laboratory frame.

	By contrast, the tin drops experience a smooth and highly symmetric driving force. The lens $\f{1}$ (with a focal length of 1\,m) forms a Gaussian beam aligned with the drop with a diffraction-limited waist $\omega_0 \sim \lLaser\, \f{1}/d_0$, where $\lambda_L=1064$\,nm is the wavelength and $d_0$ the beam diameter before lens~\f{1} \citep{hecht_optics_2002}. In our optical arrangement $\omega_0 \approx 100\,\upmu\mathrm{m}$ is much larger than the drop size $\Ro=24$ $\umu$m, which results in a homogeneous irradiation of each drop (see \Fig\ref{fig:frag:Beamprofile}\,\textit{e},~\textit{f}). Moreover, the tin drops are shielded from direct laser illumination by their own plasma cloud, which smoothes all spatial fluctuations in the laser fluence on scales smaller than $\Ro$. As a consequence, the deforming tin drops obey a high degree of rotational symmetry, as we will see in \S\ref{frag:Observation}.

\section{Phenomenology}\label{frag:Observation}
\subsection{Sequence of events for MEK drops}
\begin{figure}
  \centering
  \includegraphics{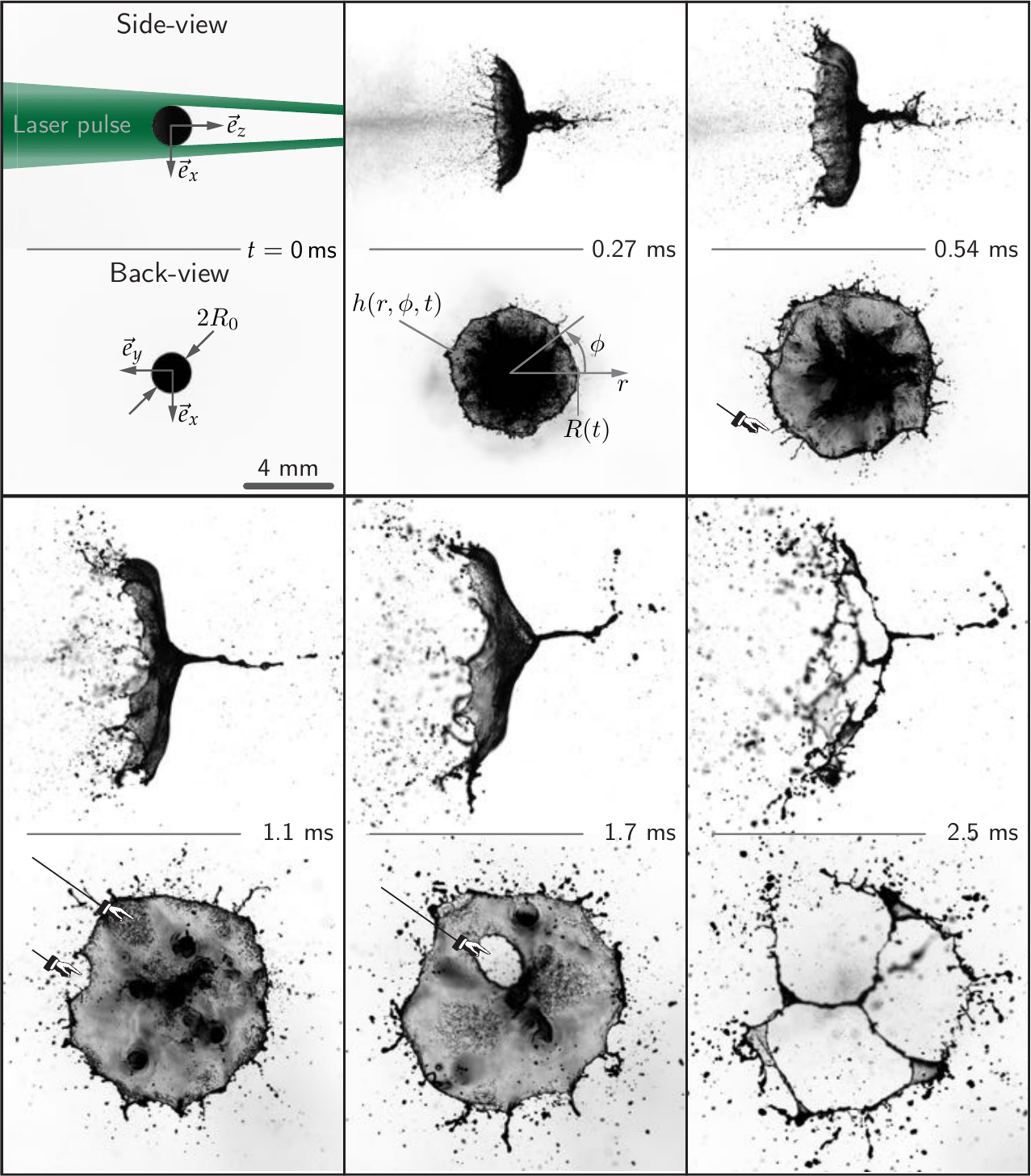}
  \caption{Sequence of events following the laser pulse impact on a MEK drop for $\We = 330$. Images are recorded stroboscopically (i.e. on different drops) from side- and back-views. The former are shown in a frame co-moving with the propulsion speed $U$. At $t =  0.27\,\mathrm{ms}$, the drop has deformed into a semi-transparent sheet with radius $R(t)$ and non-uniform thickness $h(r,\phi,t)$ that is bordered by a rim. The pointers in the three subsequent pictures indicate the onset of fragmentation of the sheet. First, \emph{rim breakup} occurs by the radial expulsion of ligaments (at $t=0.54\,\mathrm{ms}$) that subsequently destabilise. Second, corrugations of the sheet appear that finally pierce holes. This \emph{sheet breakup} occurs close to the rim, leading to \emph{neck breakup} at $t=1.1\,\mathrm{ms}$, and close to the centre of the sheet leading to \emph{centre breakup} at $t=1.7\,\mathrm{ms}$. A final web of ligaments is shown for $t=2.5\,\mathrm{ms}$.\label{fig:TimeSeries}}
\end{figure}

The MEK experiment in \Fig\ref{fig:TimeSeries} illustrates the response of a drop to the laser impact. First, the drop accelerates on the timescale~$\taue \sim 10\,\upmu\mathrm{s}$ after which it moves in the $\vec{e}_z$-direction with a velocity $U$ while it expands radially. At $t=0.27\,\mathrm{ms}$, which is close to the inertial time $\taui = \Ro/U = 0.28\,\mathrm{ms}$, the drop already resembles a thin sheet. The semi-transparent liquid reveals a thinner outer region of the sheet that is bordered by a thicker and hence darker rim. Likewise, the centre of the sheet is thick compared to the outer region. As the sheet further expands, its thickness decreases as shown by the brightening of the sheet from $t=0.54\ \mathrm{to}\ 1.7\,\mathrm{ms}$. The spatial variations of the grey level indicates that the thickness also varies in space. However, in spite of these modulations, the sheet preserves a near-circular shape during the expansion.

	While it expands, the sheet destabilises and fragments. Two types of breakup can be identified in \Fig\ref{fig:TimeSeries}. First, the breakup of the bordering rim: tiny ($\ll R$) corrugations are visible on the rim at $t=0.27\,\mathrm{ms}$ and grow over time to form ligaments (observed for the first time at $t=0.54\,\mathrm{ms}$, see pointer), which are expelled radially outward. These ligaments break into droplets that continue to move outward at a constant speed comparable to the rim velocity $\dot{R}$ at the moment of detachment. As a result of this \emph{rim breakup} at $t=1.1\,\mathrm{ms}$, the sheet is surrounded by a cloud of tiny drops.

	Second, \emph{sheet breakup} occurs through the nucleation of holes. Corrugations on the sheet are visible at $t=1.1\,\mathrm{ms}$ (a pointer at the top highlights a patch with high spatial frequency components). We observe that such disturbances on the sheet precede any hole nucleation, including events with multiple holes piercing a single patch of corrugations.
Figure~\ref{fig:TimeSeries} shows two cases where a single hole nucleates in a corrugated region. At $1.1\,\mathrm{ms}$ the lower pointer marks a hole shortly after it has pierced the sheet close to the outer rim ($r/R\sim 1$), which we term \emph{neck breakup}. At $1.7\,\mathrm{ms}$ the same process is captured in the centre of the sheet ($r/R < 0.5$, \emph{centre breakup}). 
Once a hole nucleates on the sheet it continues to grow, thereby collecting the surrounding liquid mass into ligaments. The last frame at $t=2.5\,\mathrm{ms}$ in \Fig\ref{fig:TimeSeries} shows the result of multiple holes growing and eventually merging over time. The liquid of the sheet is finally collected in a (quasi) two-dimensional structure of ligaments that breaks into droplets.

\subsection{Comparison of MEK and tin drops}\label{obs:tinvsmek}
	
\begin{figure}
  \centering
  \includegraphics{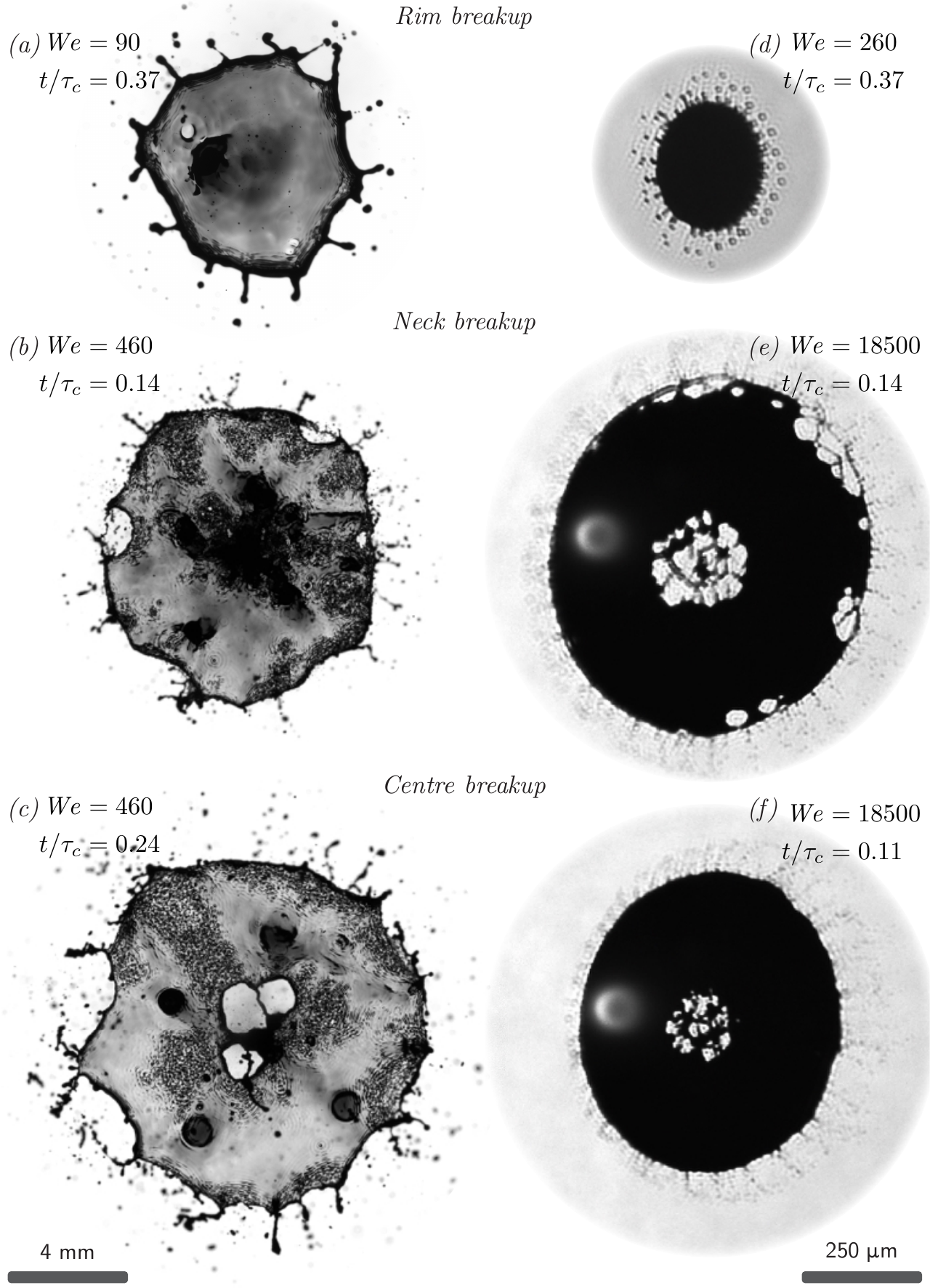}
  	\caption{Fragmentation regimes for the vapour-driven MEK drops (\textit{a}--\textit{c}, $\Ro = 0.9\,\mathrm{mm}$) and plasma-driven tin drops (\textit{d}--\textit{f}, $\Ro = 24\,\upmu\mathrm{m}$). In both systems drop fragmentation initiates at three distinct locations: the bordering rim (first row: \textit{a}, \textit{d}), the neck (second row: \textit{b}, \textit{e}) and the centre of the sheet (third row: \textit{c}, \textit{f}). The apparent elliptical shape of the tin sheets is caused by the weak parallax angle of the camera relative to the propulsion direction ($\vec{e}_z$) and is corrected for in image analysis. The white spot in \Fig\,(\textit{e}) and (\textit{f}) is an artefact of the plasma that propels the tin drops.\label{fig:FragmentationStages}}
\end{figure}

	A comparison of the fragmentation in the MEK and tin systems is presented in \Fig\ref{fig:FragmentationStages}. The first row (\textit{a}, \textit{d}) shows \emph{rim breakup} for an unpierced sheet at low Weber number. In both systems ligaments are expelled and break into droplets. In the tin sheet, the rim itself cannot be observed directly because of the tin opacity at the chosen wavelength for visualisation \citep{cisneros_dielectric_1982-1}.

	While \emph{rim breakup} is observed for MEK and tin at comparable Weber numbers, more than one order of magnitude in \We\ separates the \emph{sheet breakup} for the two systems (\Fig\ref{fig:FragmentationStages} \textit{b}, \textit{c} vs \textit{e}, \textit{f}). However, the qualitative features of the \emph{sheet breakup} are similar. In both systems the sheet breaks by the nucleation of holes in two distinct regions: \emph{neck breakup} (\textit{b}, \textit{e}) and \emph{centre breakup} (\textit{c}, \textit{f}). \emph{Neck breakup} occurs before \emph{centre breakup} and may repeat several times during the sheet expansion. 

	The observation of the \emph{neck breakup} requires a high spatial and temporal resolution. The process is strongly localised in space and difficult to separate from other breakup events. Indeed, once growing holes reach the outer rim of the sheet, the rim detaches and breaks up leaving no other trace behind than a new corrugated rim and tiny droplets. These detached drops contribute to the cloud of droplets surrounding the sheet from the \emph{rim breakup}. In \Fig\ref{fig:FragmentationStages}\,(\textit{c}) for instance \emph{neck breakup} already took place.

By contrast, the growth of holes during the \emph{centre breakup} is much easier to observe experimentally. In both MEK and tin sheets holes nucleate in the centre of the sheet, merge and collect mass in a web of ligaments that breaks up into droplets.
The opaque tin sheets prevent a further comparison of the two systems in terms of the corrugations that are visible for MEK in \Fig\ref{fig:FragmentationStages}\,(\textit{b}, \textit{c}).

\subsection{Some comments on jetting}\label{sec:jetting}

\begin{figure}
  \centering
  \includegraphics{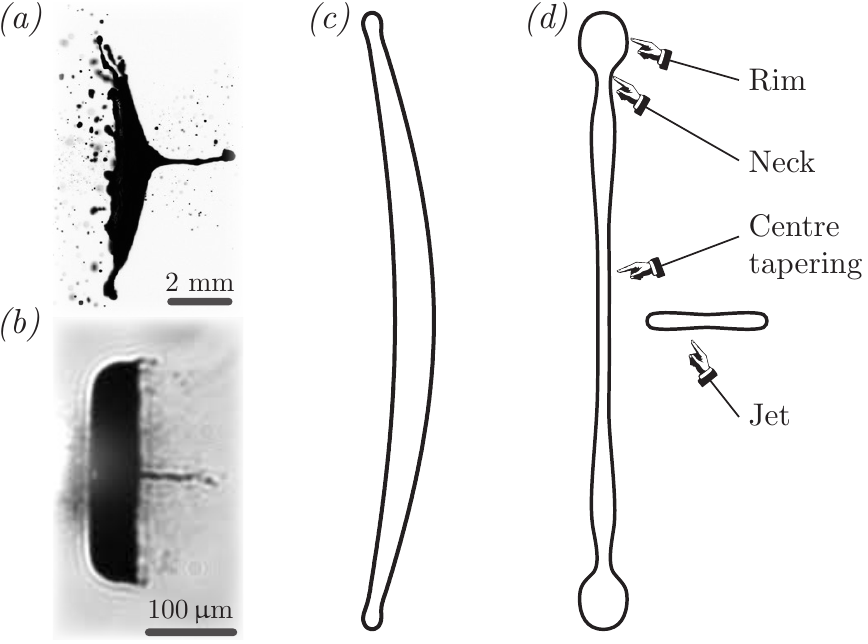}
  \caption{(\textit{a},~\textit{b}) Side-view images showing the formation of a jet in the centre of the drop in the MEK (\textit{a}) and tin (\textit{b}) systems.
  (\textit{c}) Sheet contour obtained from a boundary integral simulation illustrating the cross section of the axisymmetric shape for $\We=790$ (adapted from~\citet{gelderblom_drop_2016}).
  (\textit{d}) Sketch of the sheet showing the bordering rim and the tampered neck and centre regions.\label{fig:BaseShapeRadial}}
\end{figure}

	In addition to the \emph{rim} and \emph{sheet breakups}, one observes the ejection of mass on the opposite side of the laser impact in the form of a liquid crown (see \Fig\ref{fig:TimeSeries}). This ejected mass moves at a speed larger than $U$, collapses on the $\vec{e}_z$-axis ($t=0.54\,\mathrm{ms}$) to form a jet that detaches from the sheet and finally breaks up ($t=1.1~\mathrm{to}~2.5\,\mathrm{ms}$). A similar jetting is observed in the tin system, as shown in \Fig\ref{fig:BaseShapeRadial}\,(\textit{a}, \textit{b}). 

	This early jetting is not a direct consequence of the pressure pulse driving the drop expansion. Boundary integral (BI) simulations of the drop-shape evolution after pressure-pulse impact \citep{gelderblom_drop_2016}, which are capable to reproduce jetting phenomena in principle \citep{peters_highly_2013}, do not show this feature (see \Fig\ref{fig:BaseShapeRadial}\,\textit{c}). 

	Fast jetting often results from the implosion of a cavitation bubble \citep{crum_surface_1979,ohl_surface_2006,thoroddsen_spray_2009, utsunomiya_laser_2010, tagawa_highly_2012, gonzalez_avila_fragmentation_2016}. In the opaque tin and MEK drops ($\delta/\Ro\ll 1$) direct laser-induced cavitation is unlikely. However, pressure transients resulting from the ablation and thermoelastic effects \citep{sigrist_lasergenerated_1978,wang_thermoelastic_2001,vogel_mechanisms_2003,masnavi_simulation_2011}
and shock waves accompanying plasma generation \citep{clauer_effects_1981,marpaung_comprehensive_2001} travel through or may even focus inside the drop and induce potential cavitation spots \citep{reijers_pulses_2017}. 

	As the jet carries little mass, it has only a small effect on the overall response of the drop, and in particular on the late-time sheet dynamics. Therefore, a more detailed description of the jetting phenomenon is beyond the scope of the present study.

\section{Expansion dynamics}\label{frag:Expansion}
\subsection{Model derivation}
	
	The description of the \emph{rim} and \emph{sheet breakup} requires a model for the deformation of the drop into an expanding sheet of radius $R$ and thickness $h$. Previous models have considered a sheet with uniform thickness \citep{gelderblom_drop_2016}. However, from the MEK data it is clear that the sheet thickness has a radial dependency (see e.g. \Fig\ref{fig:TimeSeries}). Therefore, we employ here a slightly more sophisticated model that has previously been used for the sheet formed by an impact on a pillar \citep{villermaux_drop_2011}:
\begin{equation}
	\frac{R(t)-\Ro}{\Ro} = \sqrt{3\,\Wed}\, \frac{t}{\tauc}\, {\left(1-\frac{\sqrt{3}}{2} \frac{t}{\tauc}\right)}^2,\label{eq:SheetRadius}
\end{equation}
with
\begin{equation}
	\Wed = \EToEFrac \, \We, \label{eq:DefinitionWed}
\end{equation}
where $\EToE$ is the ratio of the deformation to the propulsion kinetic energies, which depends on the laser-beam profile \citep{gelderblom_drop_2016}. The rescaled Weber number $\Wed$ is only based on the fraction of the kinetic energy that is actually used for deformation. Its relation to $\We$ accounts for the difference in impact conditions between the laser case and the pillar case, as derived in Appendix \ref{sec:appendix}. 

	In the model by \citet{villermaux_drop_2011} the sheet thickness away from its axis has been described by~$h(r,t)\sim\Ro^2\Wed^{-1/2}\tauc/(rt)$, which has been validated experimentally by \citet{vernay_free_2015}. For the evolution of the sheet thickness in the centre region, which is required for the discussion on the sheet breakup in \S\ref{frag:SheetBreakup}, we use here a mass-averaged description, simply reflecting the conservation of mass,
\begin{equation}
	\frac{h}{\Ro} \sim  {\left(\frac{R}{\Ro}\right)}^{-2}.\label{eq:scalingh}
\end{equation}
 
	The energy partition $\EToE$ differs between the MEK and tin cases. In the MEK system, the relative fluence $f$ in the impact plane is kept constant for all experiments and is directly related to the recoil pressure $\pe$ as expressed by (\ref{eq:PropulsionPressure}). For the flat fluence profile observed experimentally, the energy partition can be obtained analytically  \citep{gelderblom_drop_2016}, which yields $\EToE=1.8$, independently of \ELaser.
 
\begin{figure}
	\centering
	\includegraphics{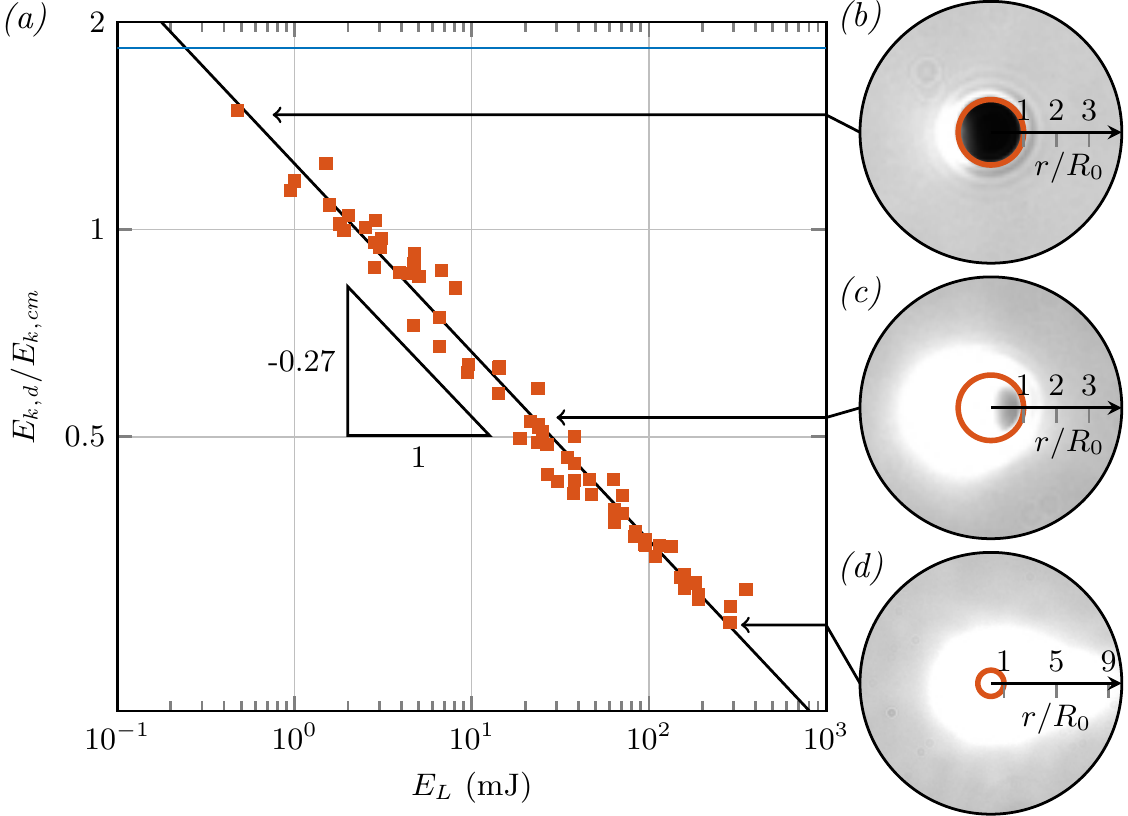}
	\caption{(\textit{a}) Energy partition as a function of laser energy for MEK (blue solid line) and tin drops (red square markers). For MEK $\EToE=1.8$, independently of $\ELaser$ as calculated analytically \citep{gelderblom_drop_2016}. The value for tin is determined for each experiment by the best fit of expression (\ref{eq:SheetRadius}) to the experimental curves shown in \Fig\ref{fig:ExpansionDynamics}. The black solid line is the power law $\EToE = 0.19\,{(\ELaser/E_0)}^{-0.27}$ with $E_0=1.0\,\mathrm{J}$ that follows from a linear regression. The three insets (\textit{b}--\textit{d}) show the white plasma clouds inducing the deformation of the tin drops (the initial undeformed tin drop is indicated in each inset by a red circle).\label{fig:EnergyPartition}}
\end{figure}

	By contrast, in the tin experiments we find that \EToE\ follows a power-law dependence on \ELaser (see \Fig\ref{fig:EnergyPartition}). This power law expresses the fact that the plasma dynamics and hence the corresponding recoil pressure is a function of the incident laser energy, even at constant focusing conditions. A theoretical prediction of the plasma dynamics goes beyond the scope of this study. However, the trend with the laser energy can be explained qualitatively: a comparison of \Fig[s]\ref{fig:EnergyPartition}\,(\textit{b}--\textit{d}) shows that at lower laser energy the plasma cloud covers a smaller area of the drop surface, which results in an effective focusing of the recoil pressure to a confined region. A focussed pressure pulse in turn results in a larger $\EToE$ \citep{gelderblom_drop_2016}. As a result, we expect \EToE\ to increase with decreasing laser energy~\ELaser, which is in agreement with the experimental observations in \Fig\ref{fig:EnergyPartition}.

\begin{figure}
	\includegraphics{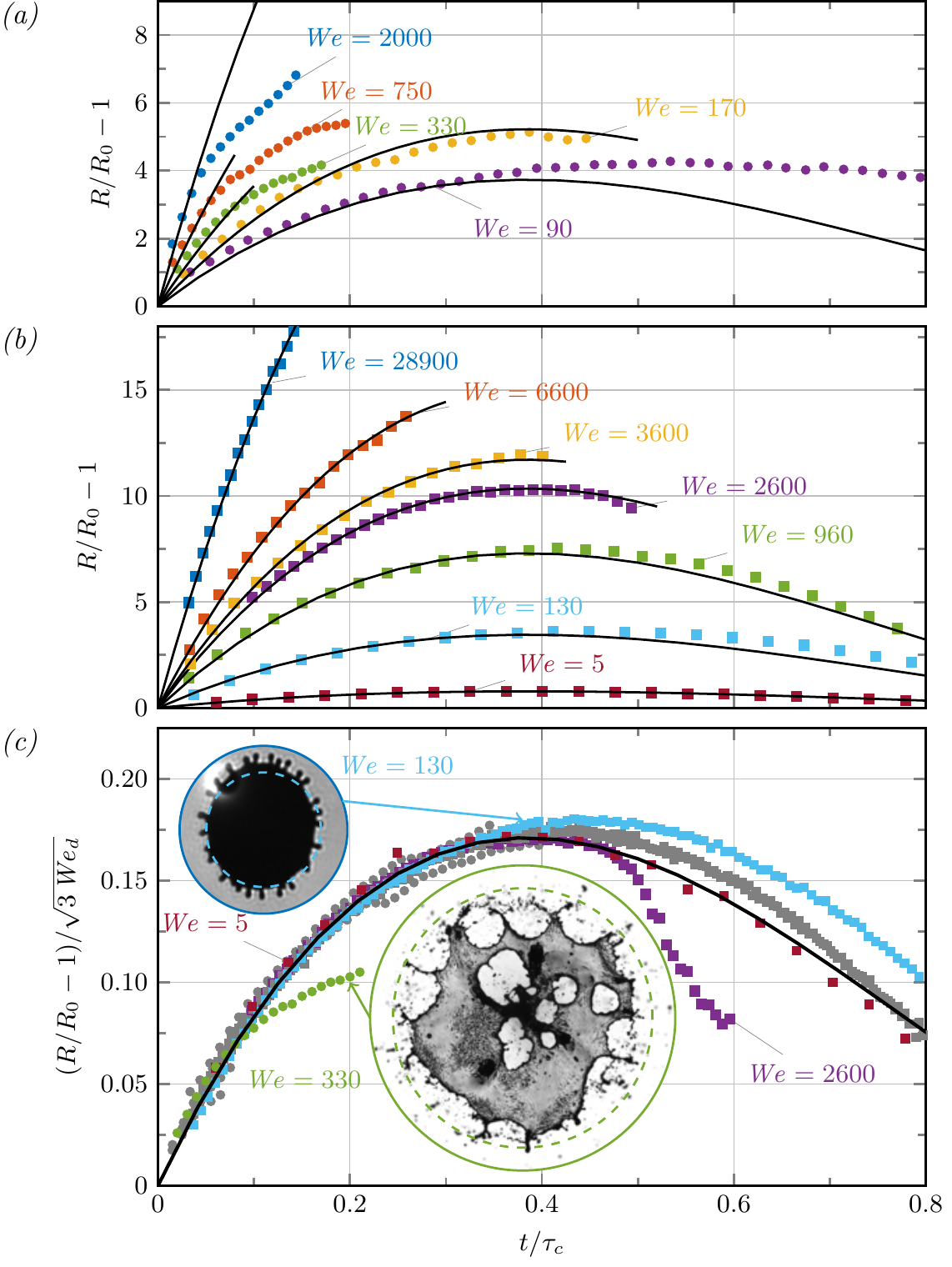}
	\caption{Sheet-radius evolution as a function of time for MEK (\textit{a}, circle markers in \textit{c}) and tin drops (\textit{b}, square markers in \textit{c}). The black solid lines represent the model \Eq(\ref{eq:SheetRadius}).	The experimental curves are shown with a reduced marker density in (\textit{a},~\textit{b}) for better visualisation. The curves ceave when the sheet evolution becomes too much affected by the fragmentation (i.e.~ when ligaments detach or holes in the sheet reach the rim). (\textit{c}) Rescaled experimental data comparing all experiments of (\textit{a},~\textit{b}) (grey markers) with the analytical prediction (\ref{eq:SheetRadius}).
	The highlighted cases and insets illustrate the influence of \emph{rim breakup} (tin drop at \textcolor{mycolor6}{$\We=130$}) and \emph{sheet breakup} (MEK drop at \textcolor{mycolor5}{$\We=330$}, tin drop at \textcolor{mycolor4}{$\We=2600$}) on the apparent sheet expansion. In the absence of fragmentation (tin drop at  \textcolor{mycolor7}{$\We=5$}) the agreement between the model and the experiments is excellent.\label{fig:ExpansionDynamics}}
\end{figure}

\subsection{Comparison between model and experiments}

	The comparison of Villermaux \& Bossa's analytical model (\ref{eq:SheetRadius}) to experiments with both MEK and tin is shown in \Fig\ref{fig:ExpansionDynamics}\,(\textit{a}) and (\textit{b}), respectively. When the experimental data are rescaled by the deformation Weber number $\Wed$ (\Fig\ref{fig:ExpansionDynamics}\,\textit{c}) they all collapse onto (\ref{eq:SheetRadius}). The model accurately captures the expansion up to the maximum radius $\Rmax$, the moment when \Rmax\ is reached at $t_\mathit{max} = 2\,\tauc/\sqrt{27} \approx 0.38\,\tauc$, and the recoil of the sheet due to surface tension.
Especially for tin the agreement between model and experiment holds over nearly four decades in Weber number (\Fig\ref{fig:ExpansionDynamics}\,\textit{b}). For MEK (\Fig\ref{fig:ExpansionDynamics}\,\textit{a}) the deviation between the model and the experimental data is larger, in particular at higher Weber numbers ($\We>170$). As we will discuss below, the model deviates from the experimental results when the fragmentation severely affects the topology of the sheet. 

	In the collapsed view of \Fig\ref{fig:ExpansionDynamics}\,(\textit{c}) a few cases are highlighted to illustrate how fragmentation affects the comparison between model and experiment. In the absence of fragmentation the experimental data follows the model closely (e.g. for tin at $\We=5$). At $\We=130$ (\refl{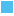}) the sheet is subject to \emph{rim breakup}.
The ligaments, which are expelled outward, do not follow the recoil and lead to an apparent over-expansion of the sheet for $t>t_\mathit{max}$ (see inset in \Fig\ref{fig:ExpansionDynamics}\,\textit{c}) since our image analysis for $R$ excludes detached ligaments but not those connected to the sheet. The same behaviour is observed for MEK at $\We=90$ (\refl{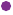}) (see \Fig\ref{fig:ExpansionDynamics}\,\textit{a}). Interestingly, the effect of the rim breakup on the sheet dynamics decreases with increasing Weber number. For $\We=960$ (\refl{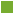}) the apparent over-expansion during the recoil phase is much smaller (\Fig\ref{fig:ExpansionDynamics}\,\textit{b}), although \emph{rim breakup} is observed in the experiments. Indeed, the sheet model (\ref{eq:SheetRadius}) predicts the rim diameter~$b$ and hence the mass contained by the rim to decrease with Weber number as $b/\Ro \sim \Wed^{-1/4}$~\citep{villermaux_drop_2011}. 

	As the Weber number is further increased, \emph{sheet breakup} in the neck region leads to a deviation between model and experiment, which is illustrated for MEK at $\We=330$ (\refl{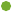}) in \Fig\ref{fig:ExpansionDynamics}\,(\textit{c}). When holes nucleating in the neck region reach the outer rim, the latter partially detaches from the sheet and the measured radius $R$ decreases rapidly (see inset). This decrease in $R$ due to the neck breakup is also visible for tin sheets, e.g.\ for $\We=2600$ (\refl{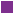}) in \Fig\ref{fig:ExpansionDynamics}\,(\textit{c}). 
The onset of the \emph{sheet breakup} occurs earlier for MEK than for tin as we will show in \S\ref{frag:SheetBreakup}. Consequently, in \Fig\ref{fig:ExpansionDynamics}\,(\textit{a}) the MEK data deviates from relation (\ref{eq:SheetRadius}) at earlier times than tin, especially for large Weber numbers where a severe \emph{neck breakup} is observed.

\section{\emph{Rim breakup}}\label{frag:RimBreakup}
\subsection{Observations}
	
\begin{figure}
	\centering
	\includegraphics{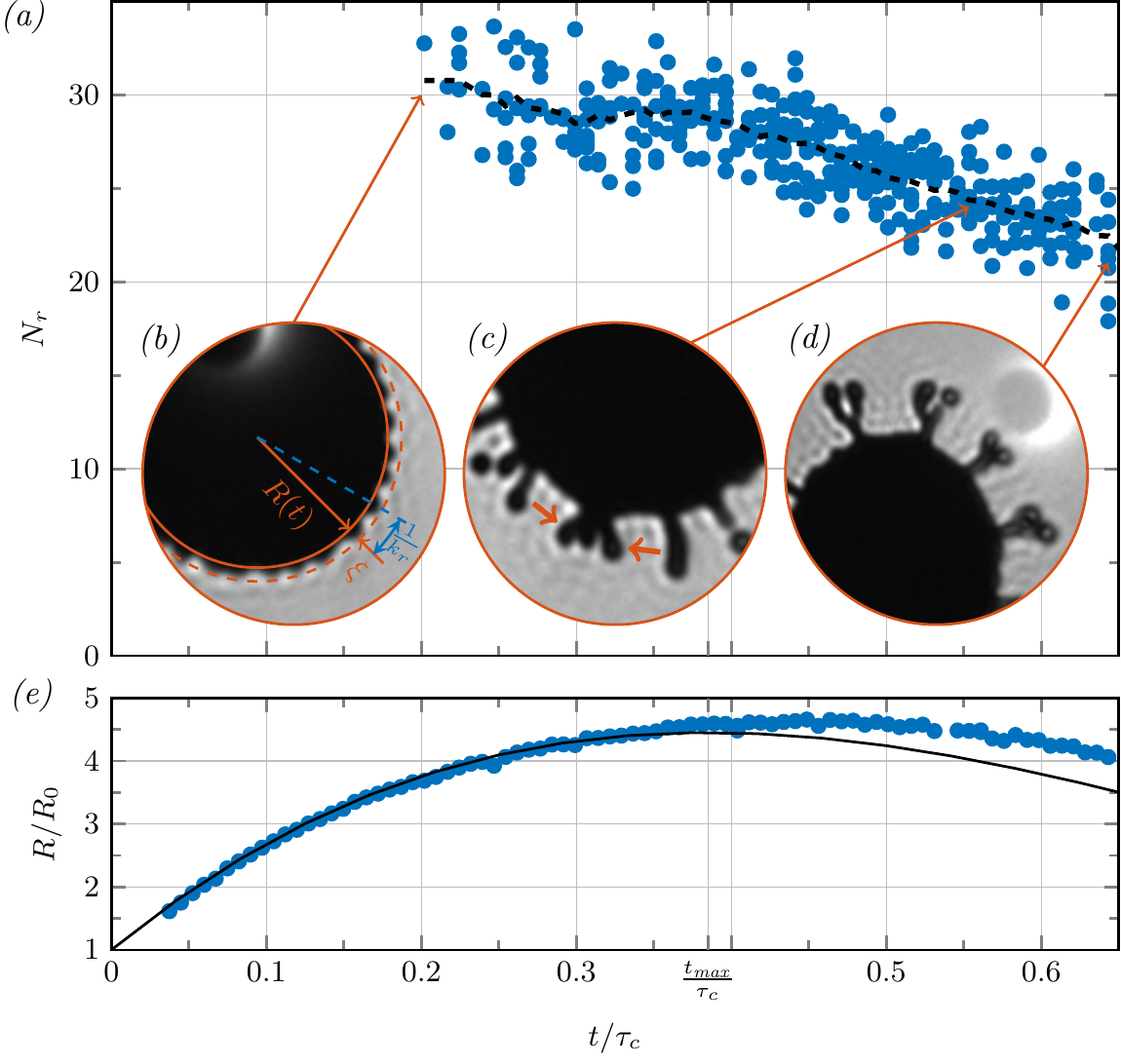} 
	\caption{Evolution of the rim breakup for $\We=132$ with the dimensionless time $t/\tauc$ obtained from a tin experiment  exhibiting a highly symmetric expansion. (\textit{a}) Total number of ligaments $\Nrl$. Each marker~(\refl{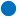}) indicates a new realisation of the experiment (with a delayed measurement) and the black dashed line~(\refl{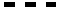}) is a running average. The inset (\textit{b}) shows the sheet radius $R(t)$, the amplitude~$\xi$, and the wavenumber~$\kl$ of the corrugation as observed at $t/\tauc=0.2$. During the recoil of the sheet ($t>\tmax$) two or more ligaments may merge as shown in insets (\textit{c}) and (\textit{d}). (\textit{e}) Sheet radius evolution. Measurements (\refl{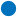}) and model \Eq(\ref{eq:SheetRadius})~(\refl{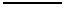}).\label{fig:WhenDoesItBreak}}
\end{figure}

	A typical evolution of the \emph{rim breakup} is illustrated in \Fig\ref{fig:WhenDoesItBreak} for tin drops with $\We=132$. Corrugations with an amplitude~$\xi$ develop on the rim. Initially, these corrugations are visible in the experiments as mere noise. Later they form clear perturbations with a characteristic wavenumber~$\kl$ from which ligaments evolve. We define the latter moment as the time~$\taul$ of \emph{rim breakup}, whereas the number $\Nrl$ of ligaments is obtained by counting.
	
\begin{figure}
	\centering
	\includegraphics{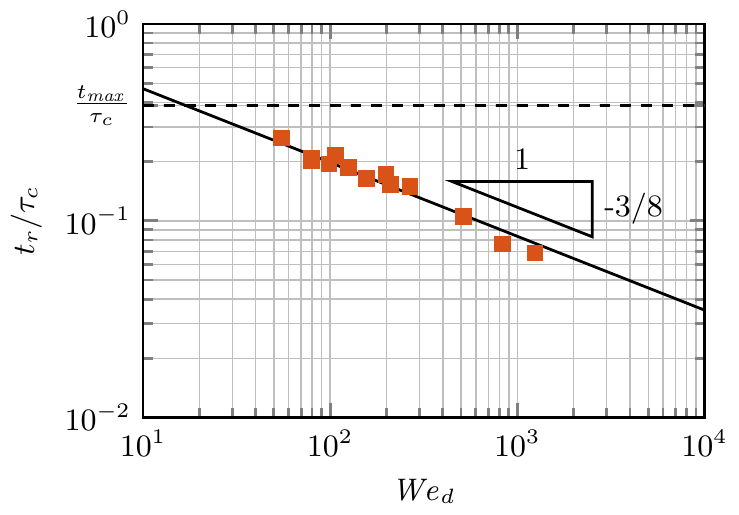}
	\caption{Time~$\taul$ when the rim corrugations become visible (see \Fig\ref{fig:WhenDoesItBreak}\,\textit{b}) as a function of the Weber number \Wed. The data is acquired manually from a subset of tin experiments that are recorded at identical camera and lighting settings to exclude any influence of the image resolution. The solid line is the scaling law (\ref{eq:taul}) with a prefactor of $1.1$.\label{fig:RimBreakupTL}}
\end{figure}

\begin{figure}
	\centering
    \includegraphics{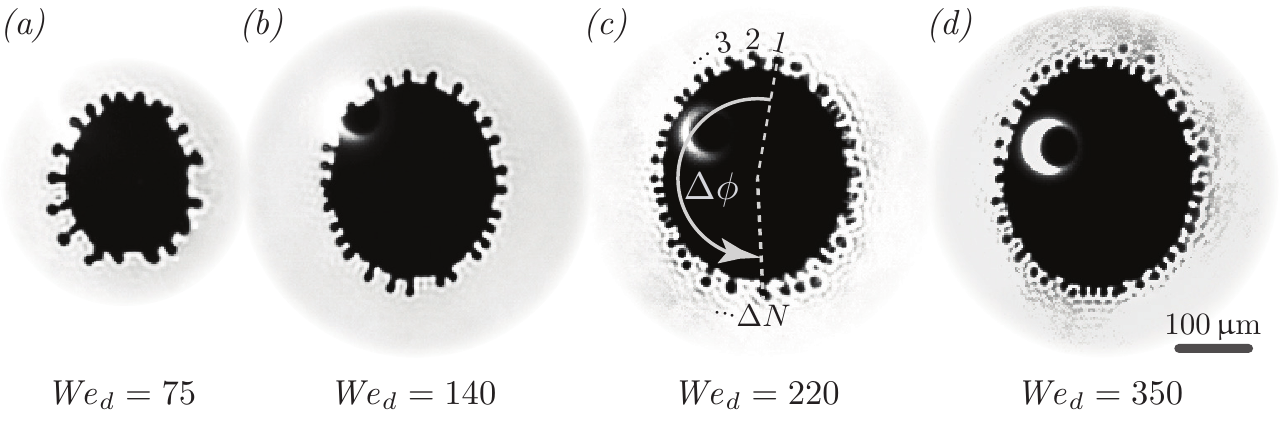}\\
	\vspace*{1em}
	\includegraphics{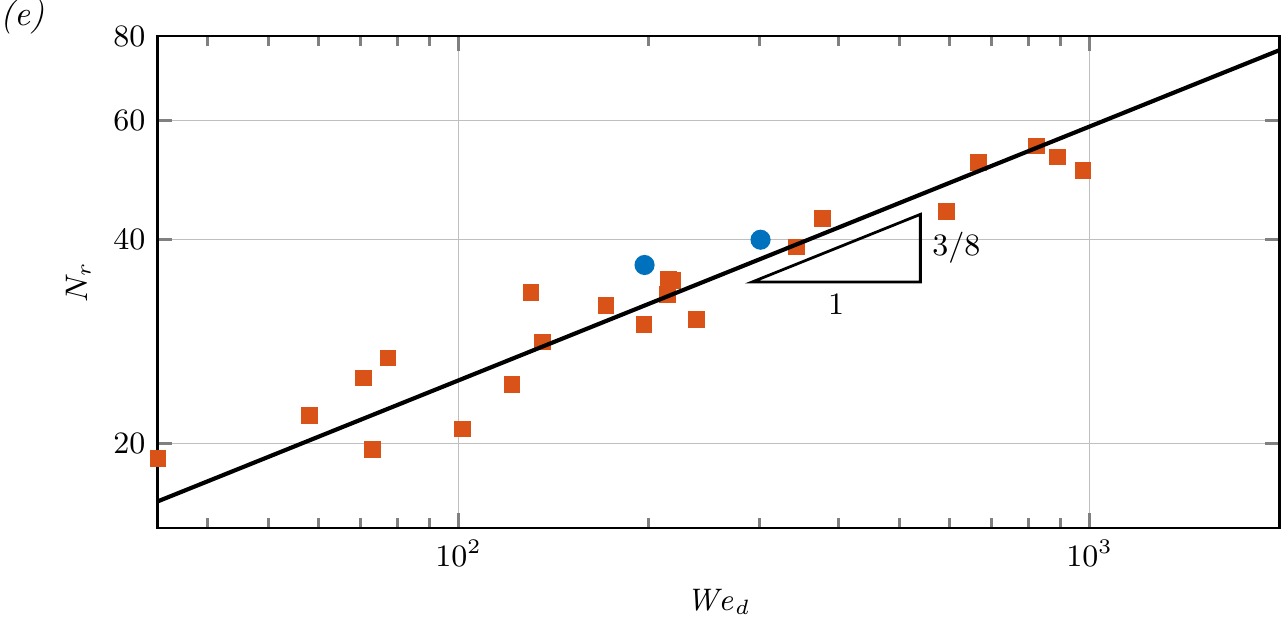}
	\caption{(\textit{a}--\textit{d}) Radial expulsion of ligaments during rim breakup for increasing Weber numbers (left to right). The back-view images are taken from experiments with tin drops exhibiting a highly symmetric expansion. When the depth of focus limits the detection of ligaments to a fraction $\Delta \phi/2\pi$ of the rim (see \textit{c}) the total number of ligaments is estimated from $\Nrl = 2\,\pi\, (\Delta N-1)/\Delta\phi$. (\textit{e}) $\Nrl$ as function of \Wed\ for tin~(\refl{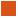}) and MEK drops~(\refl{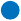}). The data for MEK is limited to two experiments since the early hole nucleation in the neck region prevents an accurate measurement of the \emph{rim breakup} for larger Weber numbers. The solid line is (\ref{eq:scalingRim}) with a prefactor of $4.4$.\label{fig:RimFragmentation}}
\end{figure}

	Figure \ref{fig:WhenDoesItBreak}\,(\textit{a}) shows that $\Nrl$ is initially constant but decreases for $t>\tmax$ due to the compression of the rim during the recoil of the sheet. These ligaments that are still attached to the sheet get closer to each other and merge from their base, as shown in \Fig\ref{fig:WhenDoesItBreak}\,(\textit{c},~\textit{d}). The rim breakup time $\taul$ is plotted in \Fig\ref{fig:RimBreakupTL} as a function of $\Wed$. Ligaments form earlier for larger Weber numbers and always form before the sheet starts retracting ($\taul<t_{max}$). The maximum number of ligaments observed over $t \le \tmax$ is found to increase with increasing \Wed\, as illustrated in \Fig\ref{fig:RimFragmentation}\,(\textit{a}--\textit{d}) with tin. This observation is confirmed by plotting $N_r$ versus $\Wed$ in \Fig\ref{fig:RimFragmentation}\,(\textit{e}). For MEK drops \emph{neck breakup} takes place much earlier than for tin and interacts with the formation of the rim ligaments. Therefore \emph{neck breakup} in MEK drops limits the range in $\Wed$ for which reliable measurement of $\Nrl$ can be obtained. However, the two measurements we obtained are in quantitative agreement with the tin data at the same Weber number.

\subsection{Model derivation and comparison with experiments}

	Inspired by the similarity with the sheet dynamics following the impact on a pillar, we follow the approach of \citet{villermaux_drop_2011} to describe the \emph{rim breakup}. We model the rim as a planar liquid cylinder of diameter $b \sim \Ro \Wed^{-1/4}$, which is justified since $\kl R \gg 1$ such that the curvature of the rim is negligible. The rim is subject to two destabilisation mechanisms. First, the Rayleigh-Plateau instability leads to a destabilisation of the rim on a timescale ${(\rho\, b^3/\gamma)}^{1/2} \sim \tauc \Wed^{-3/8}$ \citep{villermaux_drop_2011}, which agrees with our experimental observation in \Fig\ref{fig:RimBreakupTL}. Second, the rim undergoes a time-dependent deceleration $-\ddot{R}(t)$, which induces a Rayleigh-Taylor instability with growth rate $\omega\sim (\rho (-\ddot{R})^3/\gamma)^{1/4}$, because of the rim inertia.

	For high Weber numbers and large rim decelerations reached in our experiments the instability is expected and found to develop at early times ($\taul\ll \tmax$, see \Fig\ref{fig:RimBreakupTL}), in contrast to the experiments of \citet{villermaux_drop_2011}. Using $\taul\ll \tauc$, the expansion of (\ref{eq:SheetRadius}) into $\ddot{R}\sim -\Ro \Wed^{1/2}/\,\tauc^2$ gives the following timescale for the Rayleigh-Taylor instability
\begin{equation}
	\taul \sim \tauc\,\Wed^{-3/8}.\label{eq:taul}
\end{equation}
This timescale is identical  to that of the Rayleigh-Plateau instability, as already observed for liquid sheet edges in a different context by \citet{lhuissier_2011}. Figure~\ref{fig:RimBreakupTL} shows that (\ref{eq:taul}) is in excellent agreement with the experimental data with a prefactor of 1.1. The scaling (\ref{eq:taul}) differs from the breakup time $\sim \tauc$ proposed by \citet{villermaux_drop_2011} assuming that the stretching of the sheet delays the \emph {rim breakup}. 

	The sheet radius at $\taul$ and the characteristic wavenumber~$\kl$ at that time determine the number of ligaments according to  $\Nrl \sim R(\taul)\,\kl$. The fastest growing Rayleigh-Taylor mode is given by $\kl\sim {(-\ddot{R}\,\rho/\gamma)}^{1/2} \sim \Wed^{1/4}/\Ro$, identical to the characteristic wavenumber of the Rayleigh-Plateau instability. Using again the early-time expansion of (\ref{eq:SheetRadius}) we find $R/\Ro \sim \Wed^{1/2}\, t/\tauc$, which leads to 
\begin{equation}
	\Nrl \sim R(\taul)\,\kl  \sim \Wed^{3/8}.\label{eq:scalingRim}
\end{equation}
Figure \ref{fig:RimFragmentation}\,(\textit{e}) shows that (\ref{eq:scalingRim}), with a prefactor of $4.4$, is in good agreement with the tin data. Although the $\Wed$ dependence cannot be verified on the sole basis of the limited MEK data, the MEK data available is found to follow the scaling~(\ref{eq:scalingRim}) with the same prefactor as the tin data. Hence, we conclude that the difference in \emph{rim breakup} between MEK and tin is completely captured by the rescaled Weber number~\Wed\ that accounts for the different driving mechanisms, in particular the effect of the plasma dynamics on the expansion of the tin sheets.

\section{\emph{Sheet breakup}}\label{frag:SheetBreakup}

\subsection{Observations}\label{frag:SheetObservations}
 \begin{figure}
   \centering
   \includegraphics{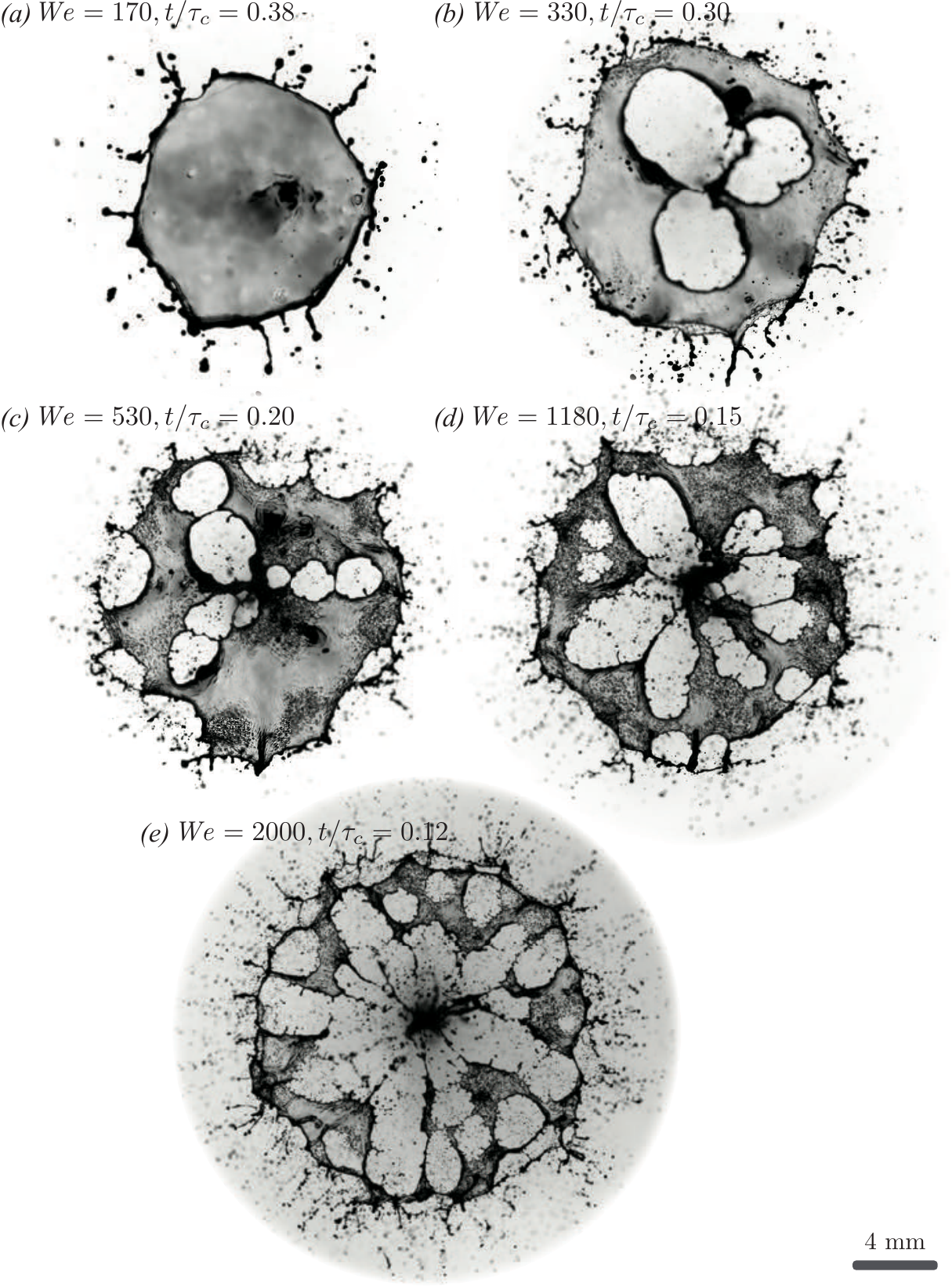}
   \caption{Sheet breakup observed from a back-view for MEK drops with increasing Weber numbers. (\textit{a}) The sheet is smooth and starts to recoil from its maximum radius $\Rmax/\Ro = 6$ reached at $\tnorm=2/\sqrt{27} \approx 0.38$, the moment the image is taken. \emph{Rim breakup} leads to the formation of ligaments but breakup of the sheet itself is not observed. A slight increase in Weber number leads to a single piercing of the sheet (not shown). (\textit{b}--\textit{e}) The sheets are pierced near their neck and in the centre before $\Rmax$ is reached. The images are taken shortly after the first centre piercing event to allow for a characteristic hole density to develop. The resulting dimensionless time of each image ($\tnorm = 0.3,\ 0.2,\ 0.15,\ 0.12$) is decreasing with increasing Weber number. The shadowgraph visualisation with a small numerical aperture is sensitive to minute light refractions and reveals the sheet corrugations just before breakup. With increasing \We\ a larger hole density resulting in a finer web of ligaments is observed at the early moment of desintegration. \label{fig:SheetFragmentation}}
 \end{figure}
  
	Figure~\ref{fig:SheetFragmentation} illustrates the \emph{sheet breakup} for MEK drops over one decade of Weber numbers. For Weber numbers up to $170$ (panel \textit{a}) the sheet remains smooth and intact at all times, and only fragments due to \emph{rim breakup}. For slightly higher Weber numbers single \emph{sheet breakup} events are observed, which are preceded by corrugations on the sheet surface (see also \S\ref{frag:Observation}). For $\We=330$ (panel \textit{b}) and higher (panel \textit{c}--\textit{e}) the sheet is more and more corrugated and ruptures both in the neck and the centre regions before it reaches its maximum expansion. The images in panels (\textit{b}--\textit{e}) are taken just after the first piercing event. They show that with increasing Weber number the \emph{sheet breakup} becomes more severe. The number of holes $\Nsl$ that pierce the sheet per unit area and the corresponding wavenumber $\kb \sim {\Nsl}^{1/2}$ increase with increasing \We. In addition, the timescale of the breakup becomes shorter as $\We$ is increased ($t/\tauc=0.30$ and 0.12 in (\textit{c}) and (\textit{e}), respectively).
\begin{figure}
	\centering
	\includegraphics{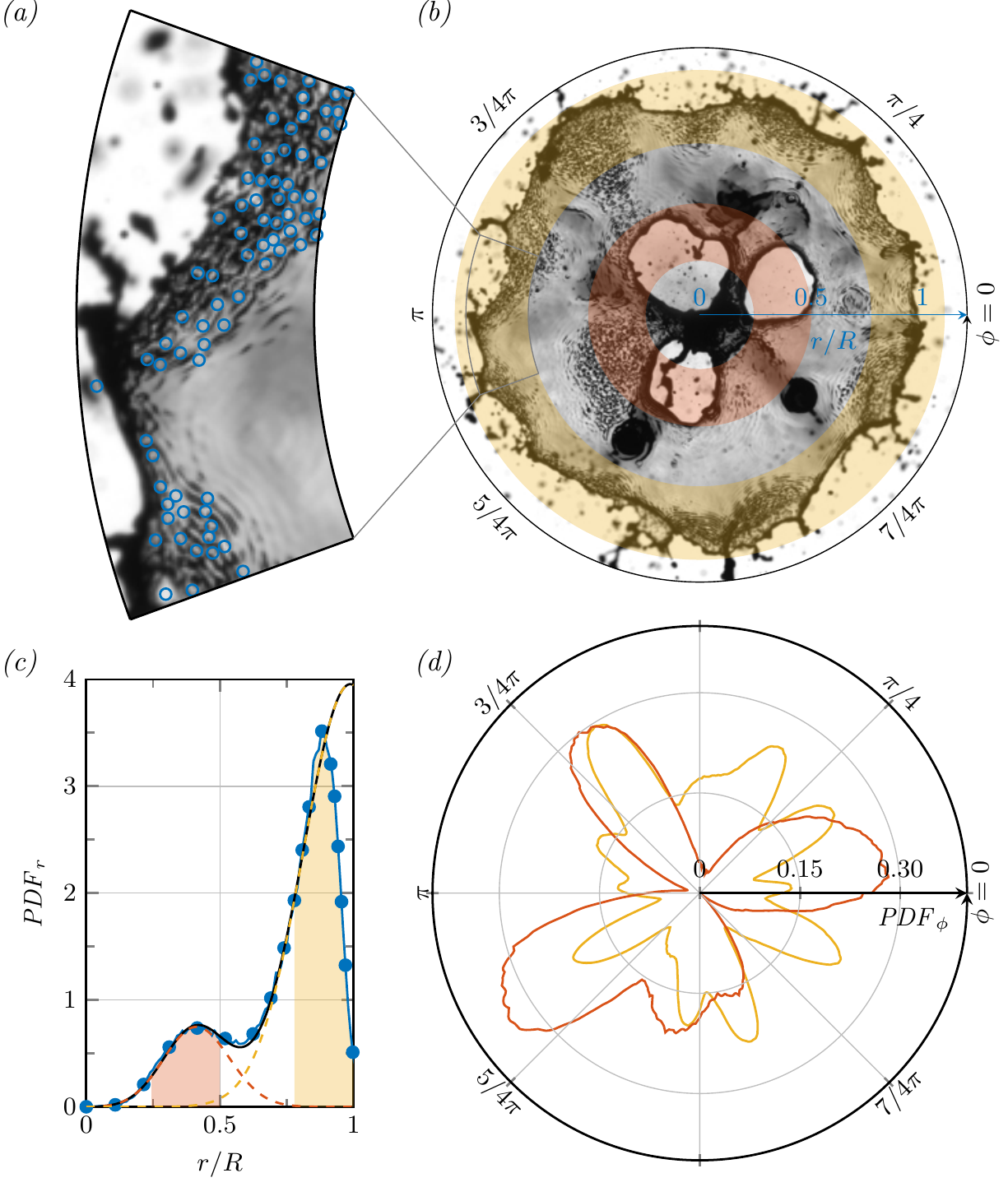}
	\caption{Corrugations and hole nucleation on MEK sheets at $\We=440$.
	(\textit{a}) Close-up view of the sheet in (\textit{b}) illustrating the result of the algorithm used to detect the corrugations that precede \emph{sheet breakup} (the local corrugations (\refl{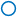}) with a spatial frequency $1/k_\mathit{corr}$ are identified by cross-correlation of the image with a gaussian kernel having a standard deviation $\sigma \sim 1/k_\mathit{corr}$). (\textit{b}) Preferred regions for \emph{sheet breakup} as identified by the analysis shown in (\textit{c}) (neck: \refl{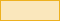}, centre: \refl{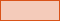}) on top of a typical sheet observed in the experiments. (\textit{c}) Probability density function (PDF) for the radial location $r/R$ of the sheet corrugations obtained from approximately 100 realisations of the experiment. The PDF is approximated by $\mathit{PDF} = 2\,r/R\ g(r)$, where $g(r)$ is a radial modulation that describes the deviation of the hole nucleation location from a \comb{spatially uniform} distribution. The experimental data (\refl{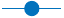}) is well described by a two-component Gaussian mixture model $g(r|\mu_i,\sigma_i)$ (\refl{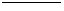}) with $\mu_i$ and $\sigma_i$ being the mean and standard deviation of the radial location of hole nucleation. The highlighted areas, i.e.\ $\mu_i - \sigma_i \le r/R \le \mu_i + \sigma_i$, illustrate the preferred hole locations in the centre (\refl{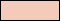}, $\mu=0.37$, $\sigma=0.13$) and the neck region (\refl{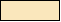}, $\mu=0.96$, $\sigma=0.18$) of the sheet. (\textit{d}) PDF of the azimuthal position $\phi$ of preferred hole locations for the centre (\refl{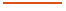}) and neck regions (\refl{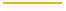}).\label{fig:ClusterFormation}}
\end{figure}

\begin{figure}
	\centering
	\includegraphics{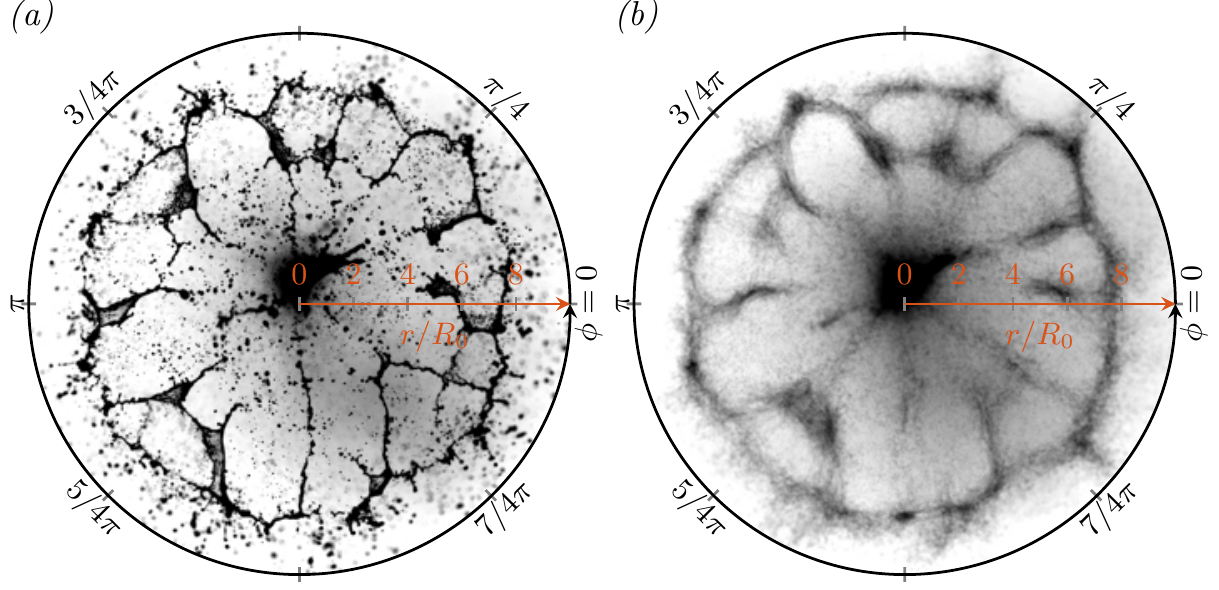}
	\caption{(\textit{a}) Back-view of a fragmented MEK sheet at $\We=2000$ and $\tnorm = 0.15$. The nucleation, growth and merger of holes on the sheet lead to a web of ligaments. (\textit{b}) Image overlay of 31 MEK sheets from 31 different drops under the same experimental condition as in (\textit{a}) and at the same time $\tnorm = 0.15 \pm 0.006$. The grey scale is proportional to the probability that a ligament be present at a given position and a black pixel means that a ligament is present at that particular position in 100$\%$ of the 31 experiments. This superposition reveals the highly deterministic nature of the final web of ligaments.\label{fig:WebOfLigaments}}
\end{figure}

	Hole nucleation in MEK is always preceded by corrugations with a high $k_\mathit{corr}$ on the sheet surface. However, no direct relation between  $k_\mathit{corr}$ and $\kb$ is found. Only a few holes pierce a corrugated area, such that $k_\mathit{corr} \gg \kb$. The corrugations can however be used as an indicator for the areas where holes are likely to nucleate. We verified this concept with an image-analysis algorithm that is sensitive to spatial frequencies much larger than the hole density, as shown in \Fig\ref{fig:ClusterFormation}\,(\textit{a}). From the data of approximately 100 experimental realisations we obtain the
probability density function (PDF) of hole nucleation in the radial direction (\Fig\ref{fig:ClusterFormation}\,\textit{c}). Not surprisingly, the quantitative analysis recovers a bimodal Gaussian distribution with two preferred areas for hole nucleation as already identified visually in \S\ref{frag:Observation}: the neck and centre region, which are marked in \Fig\ref{fig:ClusterFormation}\,(\textit{b}). For each region the PDF in the azimuthal direction is shown in \Fig\ref{fig:ClusterFormation}\,(\textit{d}). Again, there is a clear deterministic influence. Three preferred areas of hole nucleation are observed in the centre region and approximately six in the neck region. More strikingly, the final web of ligaments preserves these deterministic influences. As shown in \Fig\ref{fig:WebOfLigaments}, the web formed for a single sheet with $\We = 2000$ shows the same pattern as the overlay of 31 realisations of the same experiment. 

\begin{figure}
	\centering
	\includegraphics{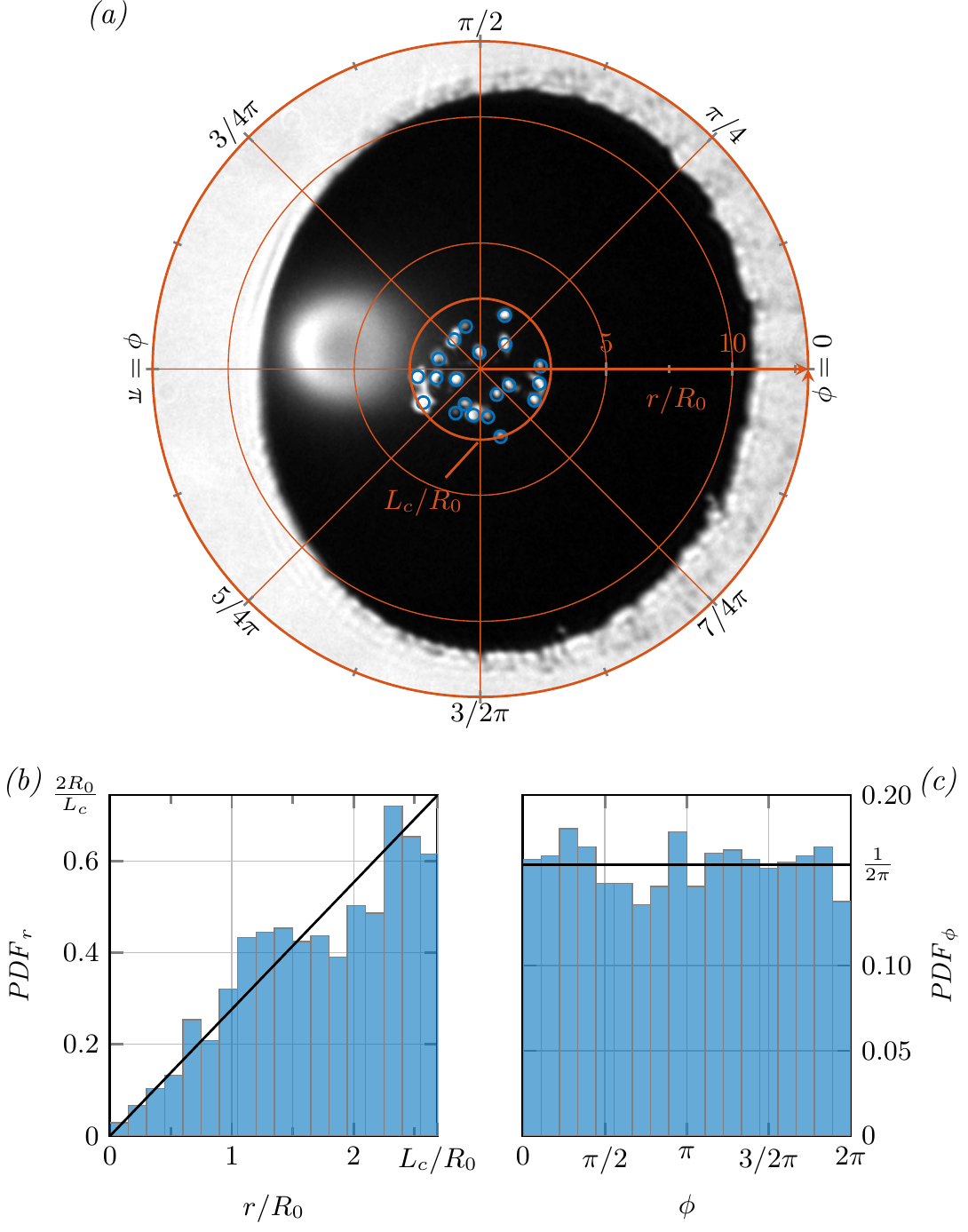}
	\caption{Hole nucleation in the centre of tin sheets at $\We=30\,000$. (\textit{a}) Tin sheet with individual holes (\refl{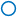}) as detected by an image-analysis algorithm sensitive to grey scale variations. The circle with radius 
$L_c =2.7\Ro$ encloses $90\,\%$ of the hole nucleation events observed over $\sim100$ realisations of the experiment. (\textit{b--c}) Radial (\textit{b}) and azimuthal (\textit{c}) distribution of nucleation events over $r\leq L_c$. The experimental distribution (\refl{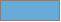}) \comb{is close to uniform}, i.e., $\mathit{PDF}_r = 2r/L_c$ and $\mathit{PDF}_\phi = 1/(2\pi)$ (\refl{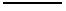}). The wavenumber of hole nucleation, $\kb \Ro =  {(\Nsl\Ro^2/(\pi n L_c^2))}^{1/2} = 0.86$, is obtained from the total number of holes~\Nsl\ observed over $n$ experimental realisations.\label{fig:HoleNucleation}}
\end{figure}

	For the opaque tin drops potential corrugations on the sheet cannot be visualised. However, as already mentioned, deterministic influences can be found in the radial location of hole nucleation by visual inspection (see \Fig\ref{fig:FragmentationStages}). In \Fig\ref{fig:HoleNucleation} we analyse the \emph{centre breakup} of a tin sheet at $\We=30\,000$. At this Weber number the hole density and radial extend $L_c$ of the centre region are such that $\kb L_c \gg 1$. Hence, we can sample a large number of holes to obtain unbiased statistics, i.e.\ unaffected by large-scale radial variations in the sheet thickness. The distribution of holes follows a linearly increasing PDF in radial direction (\Fig\ref{fig:HoleNucleation}\,\textit{b}) and uniform PDF in azimuthal direction (\Fig\ref{fig:HoleNucleation}\,\textit{c}), which express a uniform surface density in the centre region.

\subsection{Interpretation}
\subsubsection{Hole nucleation induced by a Rayleigh-Taylor instability}

	We now discuss the physical mechanism that leads to hole nucleation on the deforming MEK and tin drops. We first determine the thickness of the sheets at the moment of rupture. The minimum Weber number for sheet breakup in the MEK system ($\We\approx 170$, \Fig\ref{fig:SheetFragmentation}\,\textit{a}) translates to a radial sheet expansion of $R/\Ro \approx 6$. Similar expansions are required to observe rupture of the tin sheets. 
From the scaling relation (\ref{eq:scalingh}) this radial expansion implies a typical sheet thickness at rupture $\hb/\Ro \sim 10^{-2}$, which corresponds to an absolute sheet thickness of $\sim 10~\umu$m for MEK and $0.1~\umu$m for tin. From the high-speed recordings of individual piercing events on MEK sheets we find hole-opening speeds of $~1$\,ms$^{-1}$, which is in agreement with the Taylor--Culick speed $v = \sqrt{2\gamma/\hb\rho}\sim 1\,$ms$^{-1}$~\citep{culick_comments_1960} corresponding to our estimate of $\hb$. 
 
	From the preceding analysis we conclude that both the MEK and the tin sheets rupture when their thickness is still much larger than the length scale over which Van der Waals forces can act, which is of the order of several tens of nanometres~\citep{oron_long-scale_1997}. Furthermore, we can rule out impurities \citep{poulain_ageing_2018} as the cause of the sheet puncture. We prevent solid impurities of length scales $\sim \hb$ to enter the MEK drops by an appropriate filtration as explained in \S\ref{frag:Setup}. In the molten tin drops such large-scale impurities are also absent. From the high-speed recordings for selected MEK experiments we also exclude that the breakup is caused by individual fragments impacting on the sheet. Indeed, the ejected mass that comes from the early jetting phenomenon (see \S \ref{sec:jetting}), a likely origin for these fragments, travels at a much larger velocity than the expanding sheet and therefore cannot collide with the sheet at later times. 
 
	Hole nucleation in $\umu$m-thick, free liquid sheets has been observed by \citet{bremond_bursting_2005}. There, an impulsive acceleration of the sheet triggered a Rayleigh-Taylor instability with growing corrugations that finally pierce the sheet. 
The number of holes was found to increase with the Weber number based on the forward velocity (and hence the acceleration) of the sheet, while the characteristic rupture time decreased with \We\ \citep{bremond_bursting_2005}. 

	The sheets in our experiments are not subject to a direct acceleration of either of their interfaces. However, immediately after the laser impact the spherical drop experiences an acceleration $a \sim U/\taue = \Ro/(\tauc\ \taue)\, \We^{1/2}$ on the timescale of matter ejection $\taue$. A potential Rayleigh--Taylor instability can therefore be triggered on the drop during this early phase ($t \leq \taue$), and then develop simultaneously with the evolving sheet on the inertial timescale $\taui\sim \Ro/U$ until the sheet breaks on a timescale~$\tauc$. Since \Ro, \tauc\ and \taue\ are constant in each system, the Weber number is a direct scale for the impulsive acceleration. Experimentally, the number of holes increases and the breakup time decreases with \We\ (\Fig\ref{fig:SheetFragmentation}), as expected for the Rayleigh-Taylor sheet breakup described by \citet{bremond_bursting_2005}. Moreover, the observation that surface corrugations precede holes in the MEK sheets (figures \ref{fig:TimeSeries} and \ref{fig:ClusterFormation}) is in line with this scenario. Finally, although an instability-driven fragmentation process by itself does not explain the large scale deterministic location of the holes that is observed for both tin and MEK, we argue below that these observations are not in contradiction with an instability-induced breakup scenario.

\subsubsection{Deterministic influences on hole nucleation}\label{subsec:det}

Both MEK and tin drops show preferred spots for hole nucleation in the neck and centre regions (figures \ref{fig:FragmentationStages} and \ref{fig:ClusterFormation}). In addition, a strong deterministic influence in azimuthal direction was observed for the MEK sheets (\Fig\ref{fig:ClusterFormation}). We hypothesise that these preferred regions originate from global variations in the sheet thickness that interfere with the instability and determine where the instability can break the sheet first. These global thickness fluctuations have two different origins.
	
First, the sheet thickness is not uniform but has a thinner neck region, as was observed in the experiments with transparent MEK sheets (\Fig\ref{fig:FragmentationStages}), the sheet model (\ref{eq:SheetRadius}) and the BI simulations (\Fig\ref{fig:BaseShapeRadial}\,\textit{c}). In addition, the formation of the central jet (\Fig\ref{fig:BaseShapeRadial}\,\textit{a}, \textit{b}) induces a mass loss in the centre of the sheet. The resulting sheet thickness profile therefore has a thinner neck and centre, as illustrated in \Fig\ref{fig:BaseShapeRadial}\,(\textit{d}).
	
Second, the MEK drops are subject to an inhomogeneous laser-beam profile as explained in \S\ref{frag:BeamProfile} and shown in \Fig\ref{fig:frag:Beamprofile}\,(\textit{a}--\textit{d}). As a result, the vapour-driven MEK drops experience azimuthal modulations in recoil pressure of about $\pm 10\%$. As these modulations are deterministic, i.e.~fixed in the lab reference frame, the fragmentation also shows deterministic aspects. Azimuthal modulations are absent in the tin sheets, which result from the impact of a smooth axisymmetric laser beam (see \S\ref{frag:BeamProfile} and \Fig\ref{fig:frag:Beamprofile} \textit{e}, \textit{f}).	

\subsection{Model derivation}\label{frag:SheetModel}

\begin{figure}
  \centering
  \includegraphics{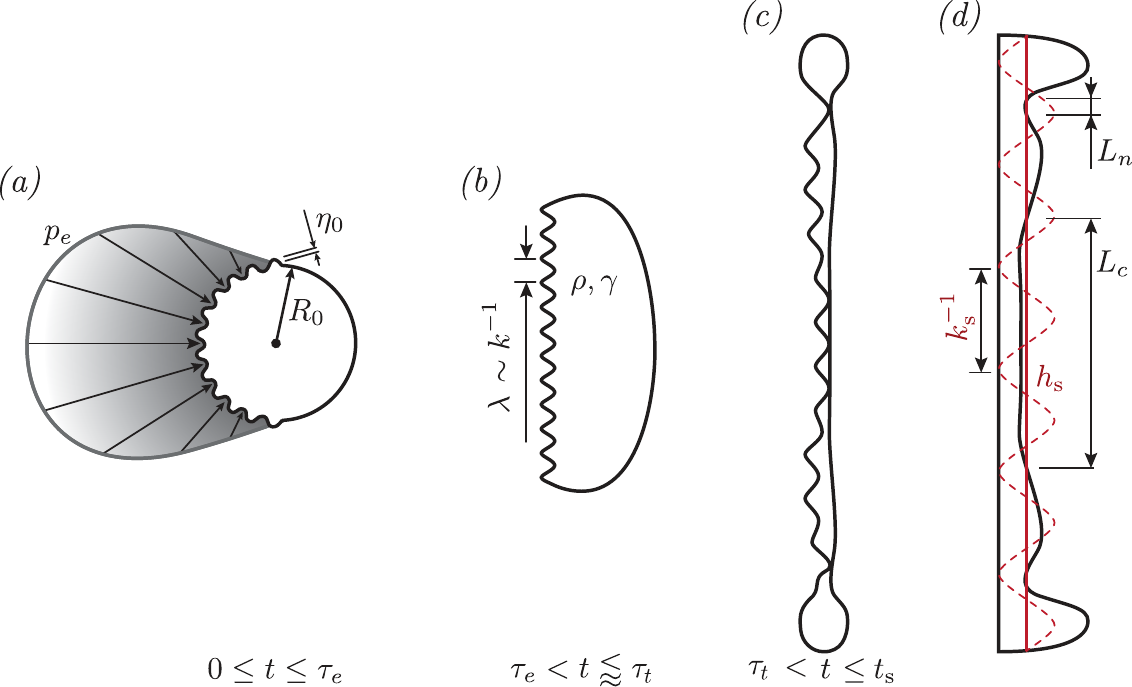}
  \caption{Sketch of the three-phase model for the evolution of the impulsive Rayleigh--Taylor instability of the deforming drop. (\textit{a}) Phase 1: the drop is accelerated perpendicular to its surface by the ablation pressure $\pe$ on timescale~$\taue$. This acceleration amplifies the Fourier modes of initial amplitude $\eta_0$ and wavenumber~$k$. (\textit{b}) Phase 2: for $\tau_e<t\lessapprox \tau_t$ the drop deforms into a sheet in the absence of any external acceleration. (\textit{c}) Phase 3: the sheet expands radially until it breaks at time $\taub$ when the perturbation amplitude is of the order of the sheet thickness~$\hb$. (\textit{d}) Detail of the sheet-thickness profile (black solid line) and the perturbation with characteristic wavenumber $\kb$ (red dashed line) that causes hole nucleation. The solid red line marks the average sheet thickness $\hb$ at the moment of breakup. In two regions where the sheet is thinnest, in the neck (marked as $L_n$) and in the centre (marked as $L_c$),  the criterion for breakup is fulfilled first and holes nucleate. \label{fig:Theory}}
\end{figure}

	We now derive a model for the Rayleigh-Taylor instability-driven \emph{sheet breakup} to obtain a prediction for the characteristic breakup time $\taub$ and wavenumber $\kb$. To this end we modify the model for \emph{sheet breakup} by \citet{bremond_bursting_2005} to account for the formation of the sheet from the spherical drop (see \Fig\ref{fig:Theory}). 
In our analysis the local thickness variations of the centre and the neck (marked by $L_c$ and $L_n$ in \Fig\ref{fig:Theory}\,\textit{d}) that lead to the preferred areas of hole nucleation discussed in \S \ref{subsec:det} are neglected. Instead, we focus on the underlying mechanism of destabilisation. Consistently, the global scaling (\ref{eq:scalingh}) is used for the overall kinematics of the sheet.

	We model the drop as a uniform sheet of initial thickness $h_0 \sim \Ro$ and density~$\rho$ that is surrounded by a gas phase of negligible density. The laser impact induces an axial acceleration of the sheet given by
\begin{equation}
 a \approx 
\begin{cases}
   \frac{U}{\tau_e}= \Ro/(\tauc\ \taue)\, \We^{1/2} &\text{for}~ 0 \le t \le \taue,\\
  0  &\text{for}~ t > \taue.\label{acc}
\end{cases}
\end{equation}
This acceleration amplifies any initial modulation of the surface, which can be represented by the Fourier modes \citep{bremond_bursting_2005}
\begin{equation}
	\eta(r,t) = \eta_0\,f(t)\,\mathrm{e}^{ikr},\label{eq:FourierGrowth}
\end{equation}
with $k$ the wave number and $r$ a generalised coordinate system tangent to the sheet. The initial amplitude $\eta_0$, which can be as small as the thermal noise in the system \citep{eggers_physics_2008}, is assumed to be characteristic to each liquid system and independent of the wavenumber. The temporal evolution~$f(t)$ follows from a potential flow analysis of the sheet and is given by $\ddot{f}(t)=-\omega^2 f(t)$, with $\ddot{f}(t)= \mathrm{d}^2f/\mathrm{d}t^2$ and $\omega(k)$ the instantaneous growth rate \citep{keller_instability_1954, bremond_bursting_2005}.

We describe the evolution of the instability on the sheet in three consecutive  phases, where we make use of the separation of timescales (\ref{eq:TimeScales_Frag}). In the first phase ($0\leq t \leq \tau_e$,~   
\Fig\ref{fig:Theory}\,\textit{a}) the drop is accelerated according to (\ref{acc}) and the modes are excited. In the second phase ($\tau_e\leq t \lessapprox \tau_t$, \Fig\ref{fig:Theory}\,\textit{b}) the acceleration is zero and the drop starts to deform. We define $\tau_t$ as the time when the transition from a deforming drop to an expanding thin sheet takes place. In \S \ref{frag:Expansion} we observed that $\tau_t\sim \Wed^{-1/2}\tau_c$, which implies that $\tau_t\sim \tau_i$. Even though during the second phase the drop no longer accelerates, the Fourier modes continue to evolve inertially, as they have acquired some velocity during the first phase. 
The third phase ($\tau_t<t\leq t_s$, \Fig\ref{fig:Theory}\,\textit{c}) is characterised by a large radial expansion $R/R_0\gg 1$ of the sheet, which stretches the Fourier modes. During this phase the sheet gets pierced at a time $t_s$ when the amplitude of the evolving perturbations equals the sheet thickness~\citep{bremond_bursting_2005}. The sheet is not uniform in thickness but has thinner regions in the neck and centre, as illustrated in \Fig\ref{fig:Theory}\,(\textit{d}). Consequently, the perturbations can pierce of the sheet in the neck and centre regions first, which thereby form preferred areas for hole nucleation. 

In the following analysis, lengths and times are non-dimensionalised by the initial drop radius \Ro\ and capillary time~$\tauc$,
\begin{equation}
	 \hat{\omega} = \omega\, \tauc,\ \hat{t}=\frac{t}{\tauc},\ \mathrm{and}\  \hat{k}=k\Ro. \label{eq:dimensionlessParameter}
\end{equation}

During phase 1 the capillary wave number $\hat{k}_c=\sqrt{\rho a \Ro^2/\gamma}= \We^{1/4} {\hat{\taue}}^{-1/2} \gg 1$ and the sheet can be considered as thick with respect to the capillary length. The dispersion relation is then given by $\hat{\omega}_1^2 = \hat{k}^3-\hat{k}_{c}^2 \hat{k}$ \citep{bremond_bursting_2005}. The modes of interest are the unstable ones that fit inside the sheet, i.e.\,$1\leq \hat{k}\leq \hat{k}_c$.
As initial conditions for the shape function $f_1(t)$ in \Eq(\ref{eq:FourierGrowth}) we assume that all modes are initially excited at the same amplitude and zero initial velocity, such that
\begin{equation}
	f_1(\hat{t}=0)=1\ \mathrm{and}\ \dot{f}_1(\hat{t}=0) = 0.\label{eq:ic}
\end{equation}
Following \citet{bremond_bursting_2005}, we treat the acceleration $a$ of the drop as impulsive, i.e.\,we assume $1/\hat{\omega}_1 \gg \hat{\taue}$. 
The sheet then behaves as an harmonic oscillator subject to an impulsive driving force, such that the shape function is given by
\begin{equation}
\ddot{f}_1(t)=-\hat{k}^3f_1(t)+\frac{We^{1/2}}{\hat{\tau}_e} \hat{k}.\label{shape1}
\end{equation}
From (\ref{eq:ic}, \ref{shape1}) we obtain
\begin{equation}
	f_1(\hat{t}) = \cos(\hat{k}^{3/2}\hat{t}) + \frac{\We^{1/2}}{\hat{k}^2 \hat{\taue}} \left\{1-\cos(\hat{k}^{3/2}\hat{t})\right\}. \label{eq:f1}
\end{equation}
To find the amplitude and growth rate of the modes at the end of phase 1, we again use the fact that the acceleration is impulsive and expand (\ref{eq:f1}) for $\hat{\tau}_e \to 0$ to obtain $f_1(\hat{\taue})\approx 1$ and $\dot{f}_1(\hat{\taue})\approx \hat{k}\We^{1/2}$. Hence, by the end of phase 1 each mode has a specific growth rate that results from the impulsive acceleration while its amplitude is still equal to unity as it did not yet have time to grow. 

	In phase 2 the modes are no longer directly amplified by an acceleration ($\hat{k}_{c}=0$) but evolve inertially. The dispersion relation therefore simplifies to $\hat{\omega}^2_2=\hat{k}^3$. The initial conditions are obtained from a matching to phase 1 at $t=\tau_e$. Again treating the acceleration as impulsive, i.e.\,letting $\hat{\tau}_e\to 0$ \citep{bremond_bursting_2005}, we find 
\begin{equation}
	f_2(\hat{t}=0) = 1\ \mathrm{and}\ \dot{f}_2(\hat{t}=0) = \hat{k} \We^{1/2}.
\end{equation}
The shape function in phase 2 is then a free harmonic oscillator
\begin{equation}
	f_2(\hat{t}) = \cos(\hat{k}^{3/2} \hat{t}) + \frac{\We^{1/2}}{\hat{k}^{1/2}} \sin(\hat{k}^{3/2} \hat{t}). \label{eq:f2}
\end{equation}

	As the drop expands into a thin sheet with $\hat{R} \gg 1$ we reach phase 3. The modes experience a stretch while at the same time the two interfaces of the sheet start to interact as their spacing $\hat{h}$ becomes of order $1/\hat{k}$, such that (\ref{eq:f2}) is no longer valid. The mode development during phase 3 is described by the thin-sheet limit of the dispersion relation ($\hat{k}\hat{h}\ll 1$) in the absence of any acceleration, $\hat{\omega}_3 = \hat{h}\hat{k}^4/2$ \citep{bremond_bursting_2005}.
The expansion of the sheet causes a self-similar stretch of the modes. At the end of phase 2 (i.e.\,just before the stretch) each mode $k$ gets deformed according to $k\Ro=k_d R$ \citep{villermaux_drop_2011}, or $\hat{k}=\hat{k}_d\hat{R}$.   
Combining this self-similar stretch in wave number with the expression for the sheet thickness (\ref{eq:scalingh}) we find for the instantaneous growth rate during phase 3: $\hat{\omega}_3\sim \hat{R}^{-6}\hat{k}^4$. Hence, while the sheet expands the growth (but also decay) rates of the modes decrease to zero much faster than the sheet thickness itself (recall equation \ref{eq:scalingh}). As a consequence, the sheet expansion and simultaneous thinning freeze the exponential growth of the modes ($\ddot{f}_3(\hat{t}) \approx 0$). The fastest growing mode $\hat{k}_\mathit{max}$ at the time $\hat{\tau}_t$ of the transition from phase 2 to phase 3 therefore determines the shape function $f_3$ according to
\begin{equation}
	f_3(\hat{t}) = f_{2,max}(\hat{\tau}_t)+\dot{f}_{2,max}(\hat{\tau}_t)\hat{t},\label{shape3}
\end{equation}  
where $f_{2,max}$ refers to the shape function (\ref{eq:f2}) evaluated for $k=k_{max}$. 
As discussed above, the transition from phase 2 to phase 3 occurs when $\hat{\tau}_t\sim\Wed^{-1/2}$. The final result of our analysis is insensitive to the prefactor in this relation, which we take equal to unity.  

	To determine $\hat{k}_\mathit{max}(\hat{\tau}_t)$ and evaluate (\ref{shape3}) we assume that the sheet expansion in phase 2 is fast in comparison with the oscillation period of $f_2$. Hence, at $\hat{\tau}_t$ the sheet is thin while all Fourier modes are still in their first oscillation period. This condition requires $\hat{k}^{3/2}\, \Wed^{-1/2} \ll 1$ (see equation \ref{eq:f2}), which is justified for our experiments where $\Wed \gg 1$. Therefore, we can expand (\ref{eq:f2}) in the limit $\hat{k}^{3/2} \hat{t} \to 0$ to obtain
\begin{eqnarray}
	f_2(\hat{t}) & \approx & 1 + \We^{1/2} \hat{k} \hat{t} - \tfrac{1}{2}\hat{k}^3 \hat{t}^2, \\
	\dot{f}_2(\hat{t}) & \approx & \We^{1/2} \hat{k} - \hat{k}^3 \hat{t}.
\end{eqnarray} 
The fastest growing mode $\hat{k}_\mathit{max}$ at the end of phase 2 is then obtained from $\mathrm{d}\dot{f}_2/\mathrm{d}k = 0$ and given by $\hat{k}_\mathit{max} = \We^{1/4} / {(3\hat{\tau}_t)}^{1/2}$. The shape function in phase 3 for $\We\gg 1$ then reads
\begin{equation}
	f_3(\hat{t})\sim \We \left(\EToEFrac\right)^{1/4}\hat{t}.
\end{equation}

	The time $\taub$ of \emph{sheet breakup} is reached when
\begin{equation}
	\hat{\eta}(t_s)=\hat{\eta}_0 f_3(\hattaub)= \hat{h}\left(\hattaub\right),\label{breakcrit}
\end{equation}
i.e.\,when the corrugation amplitude equals the sheet thickness \citep{bremond_bursting_2005}. Mass conservation dictates that the stretch in the wavelength due to sheet expansion is accompanied by a decrease in the corrugation amplitude $\hat{\eta}_0 = \hat{\eta}_d/\hat{h}$, where $\hat{\eta}_0=\eta_0/\Ro$. The breakup criterion (\ref{breakcrit}) then reduces to $\hat{\eta}_0f_3(\hattaub)=1$ and the breakup time (expressed dimensionally for convenience) reads
\begin{equation}
	\frac{\taub}{\tauc}\sim {\left(\frac{\eta_0}{\Ro}\right)}^{-1} \We^{-1} {\left(\EToEFrac\right)}^{-1/4}. \label{eq:taub}
\end{equation}
Hence, our analysis predicts how the breakup time depends on the initial amplitude of the perturbation (large initial perturbation means early breakup), the Weber number, which measures the initial acceleration of the drop, and the energy partitioning, which determines how fast the expanding sheet becomes thinner.

	From the breakup time we can find the wave number $\kb$ at breakup, which sets the hole density. To this end, we use again the self-similar stretch $\hatkb=\hat{k}_{\mathit{max}}/\hat{R}(\hattaub)$, with $\hat{R}(\hattaub)\sim \Wed^{1/2} \hattaub$ for $\hattaub\ll 1$ and $\Wed\gg 1$, to find (again dimensionally)
\begin{equation}
	\kb \Ro \sim \frac{\eta_0}{\Ro}\, \We.\label{eq:kb}
\end{equation}
Similar to the breakup time, we find that the characteristic wavenumber~$\kb$ depends on the initial amplitude of the perturbation and the Weber number. However, the self-similar stretch of the wavenumber causes the dependence on the expansion dynamics, and hence the energy partitioning, to vanish.

\subsection{Comparison between model and experiments}\label{sec:compsheet}

\begin{figure}
	\centering
	\includegraphics{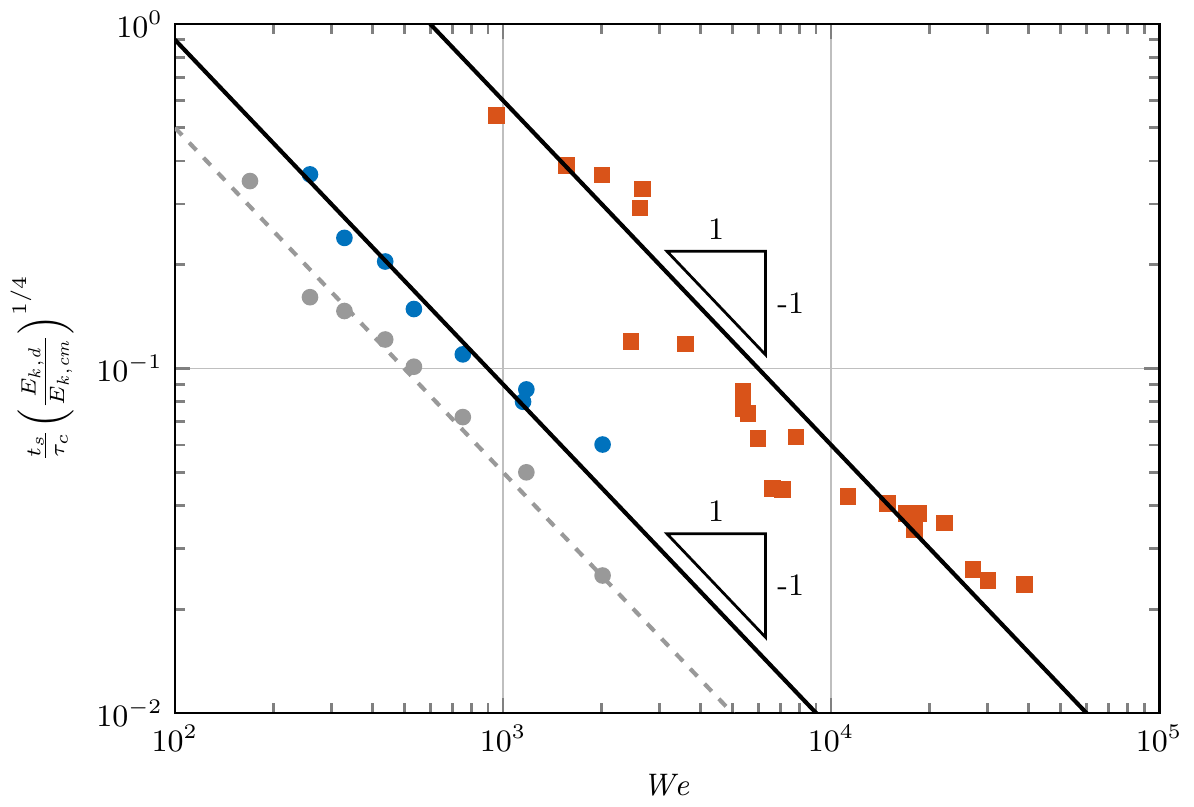}
\caption{Onset time of \emph{centre breakup} in MEK (\refl{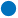}) and tin drops (\refl{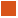}) as a function of $\We$ ($\taub$ is defined as the time at which the first hole is observed over $r/R \le 0.5$). The solid lines (\refl{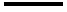}) are the prediction (\ref{eq:taub}) with a prefactor unity, $\eta_0/\Ro = 1.2 \times 10^{-2}$ for MEK and $\eta_0/\Ro = 1.3 \times 10^{-3}$ for tin. For MEK the onset time for the neck breakup is also shown in grey (\refl{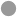}). It follows the same scaling (\ref{eq:taub}) with a prefactor of two instead of one (assuming the same noise $\eta_0/\Ro = 1.2 \times 10^{-2}$).\label{fig:BreakupTimeScale}}
\end{figure}

	Figure \ref{fig:BreakupTimeScale} compares the scaling for the breakup time (\ref{eq:taub}) with the experimental data of the \emph{centre breakup} for the tin and MEK drops. As the kinetic-energy partition \EToE\ differs between the tin and MEK experiments, the plot shows the breakup time $\taub$ rescaled by \EToE\ to allow for a direct comparison. 

	Both the MEK and the tin data sets show good agreement with the predicted $\We^{-1}$-scaling, over respectively one and two decades in Weber number. Strikingly, the absolute time at which the MEK and the tin sheets break differs by almost an order of magnitude, as was already observed in \Fig\ref{fig:FragmentationStages} in \S\ref{obs:tinvsmek}. This difference can be explained by a difference in the initial noise level from which the instability grows. Assuming a prefactor of order unity in (\ref{eq:taub}) we find for the noise level of the MEK sheets $\eta_0/\Ro=1.2 \times 10^{-2}$ and for tin $\eta_0/\Ro=1.3\times 10^{-3}$. The noise level for MEK is hence much larger than the thermal noise, which is expected to be of nanometre scale \citep{eggers_physics_2008}. Unfortunately we are not able to determine the noise level from an independent experiment. However, we can qualitatively explain the difference between MEK and tin. As discussed in \S\ref{frag:BeamProfile}, the MEK drops are subject to a much rougher beam profile and are furthermore propelled by vapour pockets bursting from their surface, while the tin drops interact with a smooth beam and plasma cloud. As a result, the initial noise in the MEK system is expected to be of macroscopic scale and much larger than for tin. These differences being accounted for by the parameter $\eta_0/\Ro$, the two data sets obtained with different liquids at vastly different length scales both confirm scaling (\ref{eq:taub}).

	In \Fig\ref{fig:BreakupTimeScale} we also show the time of destabilisation in the neck region of MEK sheets. Again, the experimental data follows the scaling (\ref{eq:taub}). Using the same noise level $\eta_0/\Ro=1.2 \times 10^{-2}$ we now obtain a prefactor of two. The different prefactor between the neck and centre region can be explained by the radial thickness profile~$h(r,t)$. For a given Weber number the neck region reaches the critical thickness $\hb$ earlier than the centre region. However, the development of the corrugation amplitude is independent of the position on the sheet: a global Rayleigh--Taylor instability is responsible for the breakup in the neck and centre region alike. Consequently, the scaling exponents for the \emph{neck} and \emph{centre breakup} are identical and in agreement with our prediction.

	Validation of the scaling for the characteristic wavenumber of breakup (\ref{eq:kb}) requires experiments with sufficient holes in the centre region, i.e.\ $\kb L_c \gg 1$, to avoid influences of the global sheet-thickness fluctuations on the statistical analysis. These conditions are out of reach for the MEK drops, whereas for tin they can only be reached at very high Weber number ($\sim 10^4$). An example of such a case  was shown in \Fig\ref{fig:HoleNucleation}. As a consequence of these extreme conditions required for statistical analysis, we were unable to experimentally validate (\ref{eq:kb}) for a broad range of Weber numbers. However, we find the order of magnitude of $\kb$ in \Fig\ref{fig:HoleNucleation} to be in agreement with (\ref{eq:kb}), assuming a prefactor of order unity and using the same noise level as obtained from \Fig\ref{fig:BreakupTimeScale}.

\section{Fragmentation regimes}\label{fragments}
\subsection{Phase diagram}
	
\begin{figure}
	\centering
	\includegraphics{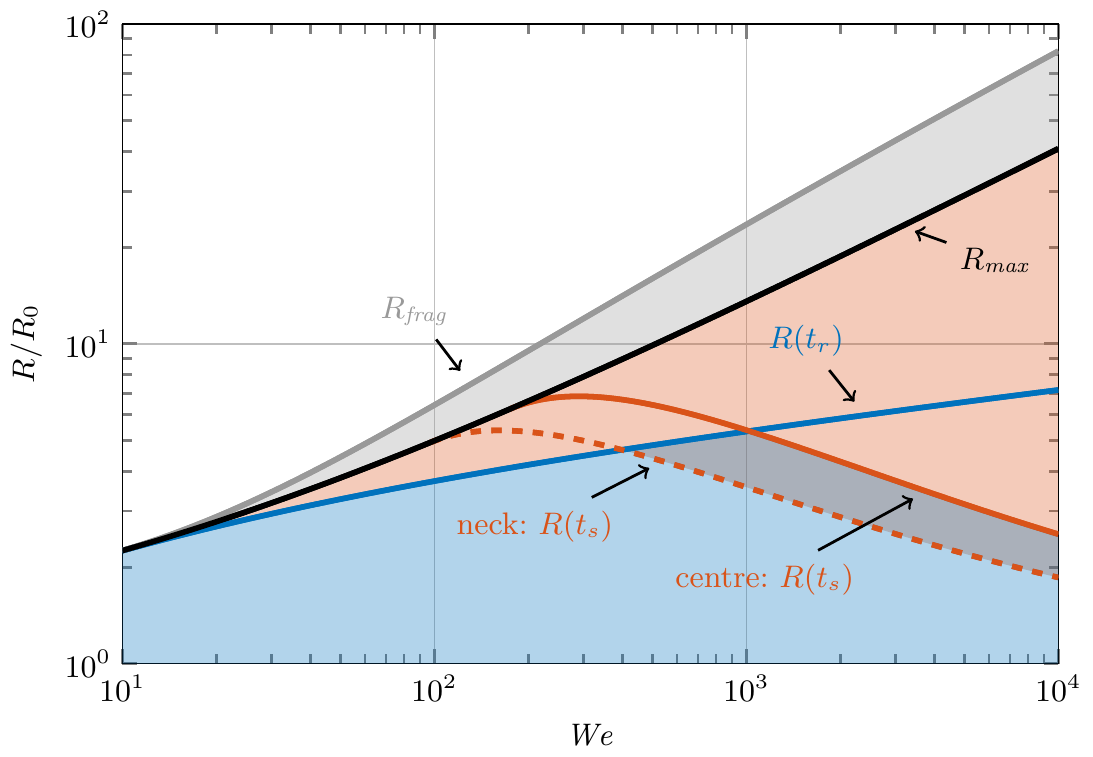}
	\caption{Overview of the drop fragmentation regimes and radial expansion in terms of the Weber number. The parameter range of stable liquid sheets (\refl{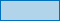}) is separated from the unstable domain (\refl{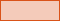}) where \emph{rim breakup} and \emph{sheet breakup} occur. The radii $R(\taul)$ and $R(\taub)$ are determined from (\ref{eq:SheetRadius}) and the scalings (\ref{eq:taul}) and (\ref{eq:taub}) for the breakup time in the rim (\refl{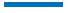}), neck (\refl{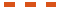}) and centre region (\refl{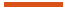}). The prefactor are those obtained from the experimental MEK data in figures \ref{fig:RimBreakupTL} and \ref{fig:BreakupTimeScale}. The maximum radius~\Rmax\ (\refl{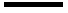}) follows from the sheet kinematics (\ref{eq:SheetRadius}) for $\tmax/\tauc = 2/\sqrt{27}$ with an energy partition for MEK of $\EToE = 1.8$. The radius~$R_\mathit{frag}$ (\refl{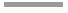}) is an estimate for the extent of the cloud of fragments (\refl{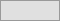}) that originate from the \emph{rim breakup} (see text).} \label{fig:regimes}
\end{figure}

	After laser impact the drop goes through a series of stages, as described above. First, the drop expands radially according to (\ref{eq:SheetRadius}). Then, at time $\taul$ given by (\ref{eq:taul}), radial ligaments evolve from the sheet rim. Finally, holes nucleate on the sheet from time $\taub$ onward given by (\ref{eq:taub}). The phase diagram in \Fig\ref{fig:regimes} summarises these different regimes as a function of the Weber number and the radial sheet expansion. The diagram is based on the scaling laws presented above with prefactors determined from the MEK experiments.

	The maximum sheet expansion $R_{max}/\Ro$ that can be achieved follows from the sheet kinematics (\ref{eq:SheetRadius}), as illustrated by the black solid line in \Fig\ref{fig:regimes}. At low Weber number this sheet remains fully intact as the accelerations of the rim and the sheet are not strong enough to trigger breakup. As the Weber number increases the rim destabilises and radial ligaments form once the sheet has reached an expansion $R(\taul)$ (blue solid line). The trajectory of fragments that originate from this \emph{rim breakup} is set by the sheet expansion rate at the moment of detachment. The fragment position at the moment of maximum sheet expansion $t_{max}$ is therefore assumed to be given by  $R_\mathit{frag}=R(\taul) + \dot{R}(\taul) (\tmax-\taul)$, as marked by the grey solid line. \emph{Sheet breakup} in the neck and centre regions occurs from $R(\taub)$ on, as marked by the red dashed and solid lines, respectively. Both breakup phenomena follow the same scaling law for the characteristic time of destabilisation $\taub$ but with a different prefactor, as discussed in \S\ref{sec:compsheet}. 

	The shaded regions in \Fig\ref{fig:regimes} indicate the different fragmentation regimes. A fully intact sheet (blue zone, \refl{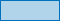}) is found at small expansion radii. A maximum intact sheet radius of $R/\Ro\approx 3.4$ is reached at $\We\approx 400$. For larger expansions the sheet always fragments. The red zone (\refl{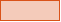}) indicates the regime where the sheet fragments, either by \emph{rim breakup} or by \emph{sheet breakup}. The radial extend of the cloud of fragments at $t_{max}$ is indicated by the grey zone (\refl{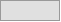}). The white zone is inaccessible due to capillary retraction of the sheet for $t>t_{max}$. 

	The phase diagram presented in \Fig\ref{fig:regimes} is a practical tool to determine the radial mass distribution of the liquid after the first laser pulse. Such information is crucial for EUV lithography applications. To access the different regimes one can either vary the Weber number by tuning the laser-pulse energy or the radial expansion of the sheet by adjusting the timing of the main laser pulse. 

\subsection{Fragment sizes}
Both the \emph{rim} and \emph{sheet breakup} give rise to a structure of elongated ligaments, which break up into droplets. The collection of all resulting fragments, which finally relax to a spherical shape, then leads to a size distribution of stable drops. 

\begin{figure}
  \centering
  \includegraphics{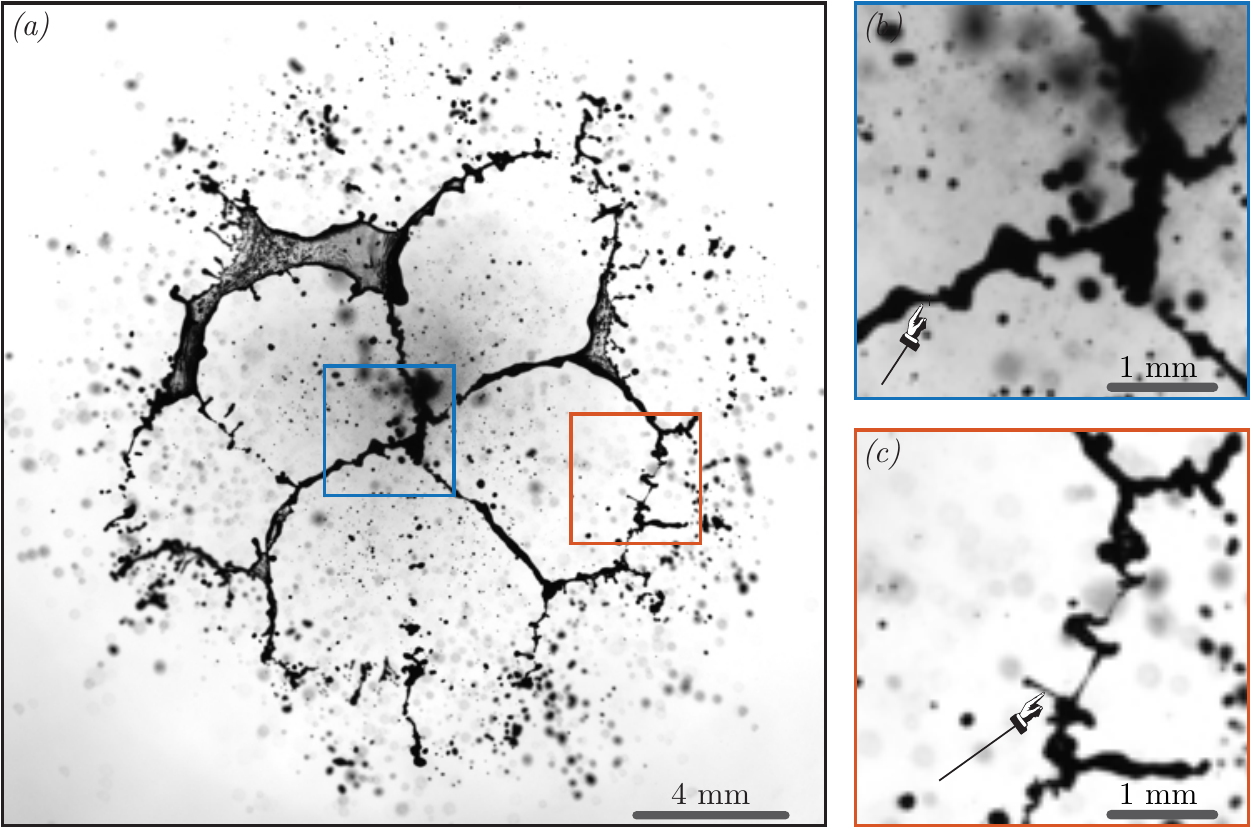}
  \caption{Web of ligaments from a MEK drop at $\We=750$ and $t/\tauc=0.27$ in (\textit{a}) and two close-up views from the centre (\textit{b}) and neck region (\textit{c}) of the sheet.
   The typical diameter of the ligaments~$d_\ell$ varies considerably between the neck and centre as exemplified by the two pointers.
   \label{fig:SizeDistribution}}
\end{figure}

\begin{figure}
    \centering
    \includegraphics{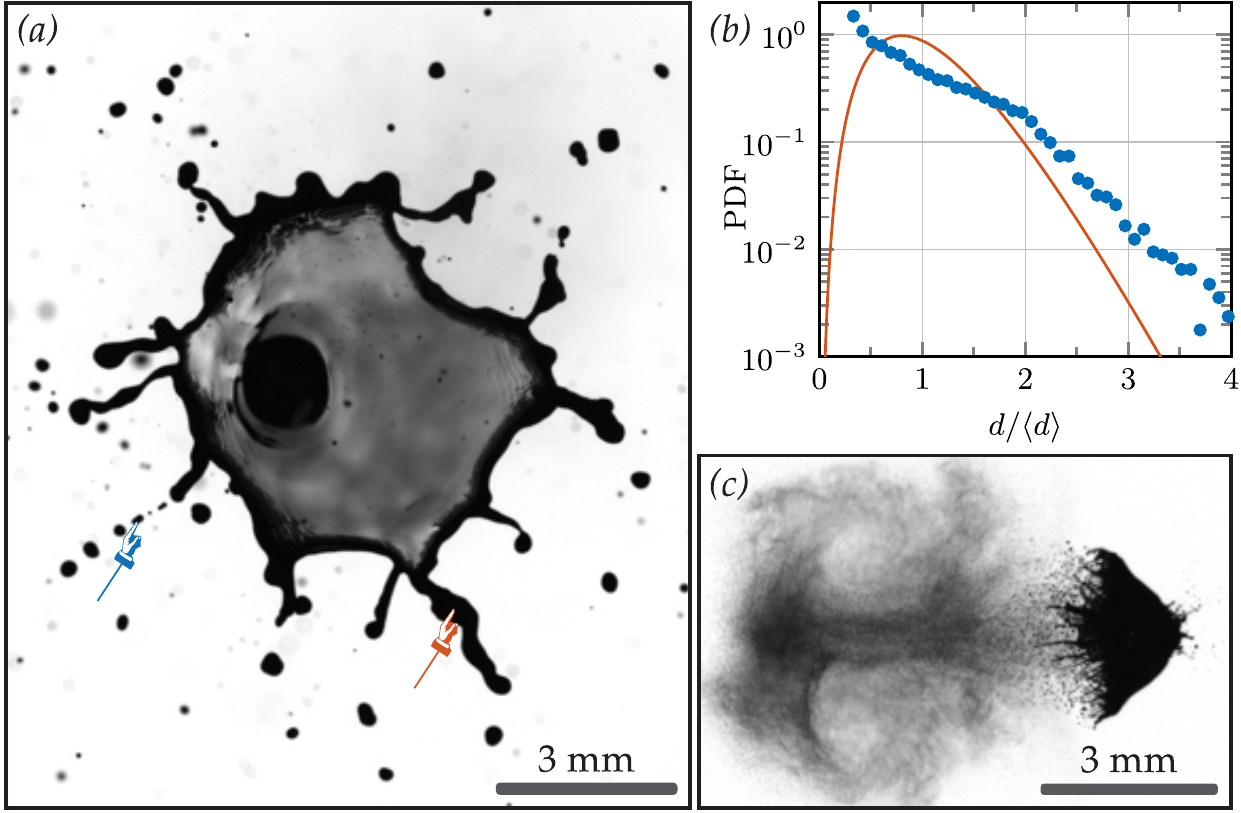}
    \caption{Fragmentation by \emph{rim breakup} at $\We=90$. (\textit{a}) The polydispersity in ligament diameter is revealed by the two highlighted ligaments with vastly different diameters ($t/\tauc=0.6$). (\textit{b}) Probability density function (PDF) of the final fragment sizes normalised by the mean fragment diameter $\langle d \rangle$. The PDF is obtained on frames similar with (\textit{a}) from 200 realisations of the experiment. The solid line is a gamma distribution of order 5 \citep{villermaux_drop_2011}. (\textit{c}) Initial mist cloud interfering with the remaining drop to generate a cloud of very small fragments ($t/\tauc=0.02$).}\label{fig:rimsize}
\end{figure}

The elementary drop-size distribution coming from a single ligament breakup depends on the roughness and the mean diameter of the ligament and can be described by a gamma distribution \citep{villermaux_fragmentation_2007}. In our experiments at least five different sources of ligaments exist, as illustrated in \Fig\ref{fig:SizeDistribution}. 
First, the rim gives rise to two types of ligaments: the radial ligaments that are expelled from the rim and the remnant of the rim itself that forms a thick circumferential ligament \citep{villermaux_drop_2011}. Second, a web of ligaments results from holes opening on the sheet \citep{lhuissier_effervescent_2013}. As the thickness of the sheet formed after laser impact is far from uniform, one might expect at least two drop size contributions originating from the neck and centre regions. 
Indeed, in \Fig\ref{fig:SizeDistribution} we observe that the mean ligament diameter varies considerably between the centre (panel \textit{b}) and neck (panel \textit{c}). Furthermore, the ligament diameter is also widely spread in each individual region (compare also \Fig\ref{fig:rimsize}\,(\textit{a}) for an example of polydisperse ligament diameters). 
A final source of very small ligaments and drops originates from the collision of rims from neighbouring holes as reported by \citet{lhuissier_effervescent_2013}. In our experiments this phenomenon is observed in particular in the neck region where the sheet is thinnest and therefore the hole-opening speed is largest. As illustrated in \Fig\ref{fig:SizeDistribution}\,(\textit{c}) rims may collide in an asymmetric fashion and form highly corrugated ligaments or even splash.
	
Because of the above mentioned, many simultaneous and sequential processes are at play in the general case, such that the analysis is {\it de-facto} challenging. Therefore, we focus on a simpler case: the fragment sizes coming from a low Weber-number impact where fragmentation occurs through the formation of radial ligaments. Figure~\ref{fig:rimsize} shows such an analysis for the MEK drops.  To obtain a PDF of the fragments sizes, we analysed 200 separate MEK experiments. At times after the rim breakup had completed, we measured the fragment sizes $d$ using a large depth-of-focus setting for the imaging equipment in order to capture all rim fragments.

Figure~\ref{fig:rimsize}\,(\textit{b}) shows the PDF of $d/\langle d \rangle$ resulting from this analysis, with $\langle d \rangle=0.15$ mm the mean fragment diameter. The PDF is cut at the optical resolution of our imaging system, which is of the order of a few $\umu$m.  Clearly, the PDF is far from the expected bell-shaped, single gamma distribution, but exhibits, in particular, too many small drops. This broad, composite size distribution presumably results from two complicating factors. First, in the MEK drops azimuthal fluctuations in the sheet thickness resulting from the inhomogeneous laser-beam profile cause the radial ligaments to have a large spread in diameter; see \Fig\ref{fig:rimsize}\,(\textit{a}). This distribution of ligaments sizes broadens the final drop size distribution. Second, there is no clear separation between fragments from the rim and from the mist cloud that originates from the drop propulsion, as illustrated in \Fig\ref{fig:rimsize}\,(\textit{c}). Hence, the many small fragments visible in the PDF do not come from the rim, but from the mist cloud. Additionally, these fast tiny mist fragments may move out of focus, causing a bias in the analysis of the smallest fragments. The interference of the mist cloud with the rim fragments also affects the mean fragment diameter $\langle d \rangle/2R_0=0.088$, which is about a factor three smaller than one would expect for pure rim fragmentation, whose fragment size distribution does not present this small-size divergence \citep{villermaux_drop_2011}.

Tin drops do not suffer from these complications. However, for these much smaller drops the analysis is severely limited by the optical resolution of the shadowgraph images. The small ligaments in \Fig\ref{fig:SizeDistribution}\,(\textit{c}) for the MEK system suggest a minimum ligament diameter of $d_\ell/\Ro \sim 10^{-3}$. In absolute terms this size would translate to a few tens of nanometres for the tin system, i.e.\ far below the resolution of optical imaging in the visible spectrum. 
  
 The qualitative analysis presented here underlines the difficulty of establishing in a non-ambiguous manner a drop size distribution arising from a non-trivial fragmentation process when the origin of each fragment cannot be traced back to a precisely identified intermediate mechanism. This is sometimes possible \citep{lhuissier_effervescent_2013,vledouts_explosive_2016} and when it is not, the analysis is often bounded to invoking general principles in lumped descriptions (see e.g. \cite{He_17} in the context of laser-pulse fragmentation, and the other examples discussed in section 7 of \cite{vledouts_explosive_2016}), a pitfall we conscientiously avoid here.
 
\section{Discussion \& conclusion}

	We have studied the fragmentation of a liquid drop that is propelled by a laser-induced phase change. Two liquid systems have been considered that differ not only in length scale but also in propulsion mechanism: millimetre-sized vapour-driven drops of methyl ethyl ketone (MEK) and micron-sized plasma-driven tin drops. These systems are subject to similar destabilisation mechanisms and have allowed us to study drop fragmentation over a wide range of Weber numbers. In addition, the two systems reveal how the early-time laser-matter interaction influences the late-time drop fragmentation.

	For both systems, fragmentation has been found to result from two Rayleigh-Taylor instabilities that are caused by accelerations of the drop in two orthogonal planes and at different timescales. First, the drop expands radially into a thin sheet with a decelerating rim from which ligaments get expelled. This \emph{rim breakup} is similar to the one encountered on liquid sheets formed after the impact of a drop onto a pillar \citep{villermaux_drop_2011}. Second, the laser impact causes an impulsive forward acceleration of the drop. As a consequence, the evolving sheet destabilises through the nucleation of holes, which we referred to as \emph{sheet breakup}. This destabilisation is similar to that of the impulsively accelerated soap film described by \citet{bremond_bursting_2005} but differs in a crucial aspect: in our experiments both the formation of the film and its destabilisation are the result of the very same initial impact. 

	The laser-matter interaction affects the drop fragmentation in several ways. First, it governs the deformation of the drop into a sheet. The resulting \emph{rim breakup} depends on two parameters that are controlled by the laser: (i) the Weber number based on the propulsion speed of the drop, which is set by the laser-induced recoil pressure acting on the drop surface and therefore depends on the laser-pulse energy, and (ii) the kinetic energy partition between expansion and propulsion. This parameter depends on the laser-beam profile, and (for tin) on the extend of the plasma cloud, which in turn depends on the pulse energy. 

	Second, the laser-matter interaction gives rise to tiny perturbations on the drop surface that grow over time and finally lead to \emph{sheet breakup}. Consistently, our analysis of the \emph{sheet breakup} involves a third parameter, besides the Weber number and energy partition, to distinguish between the different driving mechanisms: the amplitude $\eta_0/\Ro$ of the corrugations that are present during the initial acceleration of the drop. This initial noise explains the early \emph{sheet breakup} for MEK in comparison to the tin system: the noise level in the vapour-driven MEK drops is an order of magnitude larger than for the plasma-driven tin drops. 

	Third, the laser-matter interaction governs global (large scale) thickness fluctuations on the expanding sheet, both in radial and (for MEK) in azimuthal direction. These fluctuations are either a direct result of irregularities in the laser-beam profile, or originate from non-uniformities in the sheet kinematics, which in turn are determined by the laser impact. The random hole nucleation induced by the instability gets convoluted by this deterministic profile in the sheet thickness. As a result, the final web of ligaments formed by the \emph{sheet breakup} shows deterministic features. 

	Accounting for the differences in laser-matter interaction in tin and MEK  through the Weber number, energy partition and initial noise, we could explain the expansion dynamics, timescale and wavenumber for the \emph{rim breakup} and the timescale for \emph{sheet breakup} from the same model. Both types of Rayleigh-Taylor instabilities induced by the laser impact lead to the formation of ligaments, which finally break into droplets due to a Rayleigh-Plateau instability \citep{villermaux_fragmentation_2007}. 

	At least five different types of ligaments resulting from \emph{rim} and \emph{sheet breakup} were identified. The resulting drop-size distribution is very broad, even for the simplest case with only \emph{rim breakup}. While the tin fragments are too small to be characterised in a reliable way, the MEK system shows deterministic facets that are set by the laser-beam profile and not of universal nature. These influences need to be incorporated in a statistical analysis in a transparent way to obtain a description that holds for both the tin and MEK system. Such an analysis clearly deserves further investigation. The work presented here is the first step towards a full description of drop fragmentation by laser impact that incorporates both the chaotic and deterministic facets of ligament formation.

\begin{acknowledgements}
We thank Dries van Oosten, Luigi Scaccabarozzi, Jacco H.\ Snoeijer and Michel Versluis for fruitful discussions.
This work is part of an Industrial Partnership Programme
of the Netherlands Organization for Scientific Research (NWO). This
research programme is co-financed by ASML.
\end{acknowledgements}

\appendix
\section{Rescaling the late-time sheet model}\label{frag:Rescale}\label{sec:appendix}
The impact of a drop with the velocity $U$ and initial radius~\Ro\ on a pillar of the same size leads to the development of a radial sheet of thickness $h(r,t)$ and radius~$R(t)$ with dynamics~\citep{villermaux_drop_2011}
\begin{eqnarray}
	\frac{R(t)-\Ro}{\Ro} &=& \sqrt{\mathit{\tilde{We}}}\, \frac{t}{\tauc}\, {\left(1-\frac{\sqrt{3}}{2} \frac{t}{\tauc}\right)}^2, \label{app:R} \\
	h(r,t) &\sim& \frac{\Ro^3}{U r t} = \frac{\Ro^2\,\tauc}{\sqrt{\tilde{\We}}\, rt},\label{app:h} \\
	u_r(r,t) &=& \frac{r}{t}\label{app:u},
\end{eqnarray}
where $u_r$ is the radial velocity inside the sheet. Here, the rescaled Weber number $\tilde{\We}$ accounts for different initial conditions during a laser impact in comparison to a mechanical impact on a pillar. In analogy to \citet{gelderblom_drop_2016} we find $\tilde{\We}$ from a matching to the initial kinetic energy partition obtained from an early-time ($t\leq\tau_e$) model of the drop, termed $E_\mathit{k,d}/E_\mathit{k,cm}$. Matching to the energy partition in terms of sheet model (\ref{app:R}-\ref{app:u}) then reads
\begin{equation}
\EToEFrac = \frac{\int_{0}^R u_r^2 \, hr\,\mathrm{d}r}{U^2\int_{0}^R  hr\,\mathrm{d}r} = \frac{R^2}{3 U^2 t^2}, \label{app:etoemid}
\end{equation}
where $E_{k,d}$ is the kinetic energy associated with the deformation (expansion) of the sheet and $E_{k,cm}$ with the kinetic energy of the centre-mass-motion. For $t\ll \tau_c$ and $\Wed\gg 1$ we approximate (\ref{app:R}) by $R \approx \Ro \sqrt{\tilde{\We}} \,t/\tauc$ such that by using that $U^2 = \We\, \Ro^2/\tauc^2$ we find 
\begin{equation}
	\tilde{\We} = 3\, \We \EToEFrac=3\,\Wed,
\end{equation}
with \Wed\ as defined in \Eq(\ref{eq:DefinitionWed}).
This results explains our rescaling in \Eq(\ref{eq:SheetRadius}). The energy partition for the flat-top beam profile used in the MEK system is obtained analytically from the early-time model as $\EToE \approx 1.8$ \citep{gelderblom_drop_2016}.

\bibliographystyle{jfm}
\bibliography{main}

\begin{thebibliography}{43}
\expandafter\ifx\csname natexlab\endcsname\relax\def\natexlab#1{#1}\fi
\def\au#1{#1} \def\ed#1{#1} \def\yr#1{#1}\def\at#1{#1}\def\jt#1{\textit{#1}}
  \def\bt#1{#1}\def\bvol#1{\textbf{#1}} \def\vol#1{#1} \def\pg#1{#1}
  \def\publ#1{#1}\def\arxiv#1{#1}\def\org#1{#1}\def\st#1{\textit{#1}}

\bibitem[Banine {\em et~al.\/}(2011)Banine, Koshelev \&
  Swinkels]{banine_physical_2011}
{\sc \au{Banine, V.~Y.}, \au{Koshelev, K.~N.} \& \au{Swinkels, G. H. P.~M.}}
  \yr{2011}  \at{Physical processes in {{EUV}} sources for microlithography}.
  {\em\protect\JournalTitle{J. Phys. D Appl. Phys.}}  \bvol{44}~(25),
  \pg{253001}.

\bibitem[Bremond \& Villermaux(2005)]{bremond_bursting_2005}
{\sc \au{Bremond, N.} \& \au{Villermaux, E.}} \yr{2005}  \at{Bursting thin
  liquid films}. {\em\protect\JournalTitle{J. Fluid Mech.}}  \bvol{524},
  \pg{121--130}.

\bibitem[Cisneros {\em et~al.\/}(1982)Cisneros, Helman \&
  Wagner]{cisneros_dielectric_1982-1}
{\sc \au{Cisneros, G.}, \au{Helman, J.~S.} \& \au{Wagner, C. N.~J.}} \yr{1982}
  \at{Dielectric function of liquid tin between 250 and 1100$\,^\circ${{C}}}.
  {\em\protect\JournalTitle{Phys. Rev. B}}  \bvol{25}~(6),  \pg{4248--4251}.

\bibitem[Clauer {\em et~al.\/}(1981)Clauer, Holbrook \&
  Fairand]{clauer_effects_1981}
{\sc \au{Clauer, A.~H.}, \au{Holbrook, J.~H.} \& \au{Fairand, B.~P.}} \yr{1981}
   \at{Effects of laser induced shock waves on metals}.  \bt{In {\em Shock
  Waves and High-Strain-Rate Phenomena in Metals\/}},  \pg{pp. 675--702}.
  \publ{{Springer}}.

\bibitem[Crum(1979)]{crum_surface_1979}
{\sc \au{Crum, L.~A.}} \yr{1979}  \at{Surface oscillations and jet development
  in pulsating bubbles}. {\em\protect\JournalTitle{J. Phys. Colloq.}}
  \bvol{40}~(C8),  \pg{C8--285--C8--288}.

\bibitem[Culick(1960)]{culick_comments_1960}
{\sc \au{Culick, F. E.~C.}} \yr{1960}  \at{Comments on a {{Ruptured Soap
  Film}}}. {\em\protect\JournalTitle{J. Appl. Phys.}}  \bvol{31}~(6),
  \pg{1128--1129}.

\bibitem[Eggers \& Villermaux(2008)]{eggers_physics_2008}
{\sc \au{Eggers, J.} \& \au{Villermaux, E.}} \yr{2008}  \at{Physics of liquid
  jets}. {\em\protect\JournalTitle{Rep. Prog. Phys.}}  \bvol{71}~(3),
  \pg{036601}.

\bibitem[Gelderblom {\em et~al.\/}(2016)Gelderblom, Lhuissier, Klein, Bouwhuis,
  Lohse, Villermaux \& Snoeijer]{gelderblom_drop_2016}
{\sc \au{Gelderblom, H.}, \au{Lhuissier, H.}, \au{Klein, A.~L.}, \au{Bouwhuis,
  W.}, \au{Lohse, D.}, \au{Villermaux, E.} \& \au{Snoeijer, J.~H.}} \yr{2016}
  \at{Drop deformation by laser-pulse impact}. {\em\protect\JournalTitle{J.
  Fluid Mech.}}  \bvol{794},  \pg{676--699}.

\bibitem[Gonzalez~Avila \& Ohl(2016)]{gonzalez_avila_fragmentation_2016}
{\sc \au{Gonzalez~Avila, S.~R.} \& \au{Ohl, C.-D.}} \yr{2016}
  \at{Fragmentation of acoustically levitating droplets by laser-induced
  cavitation bubbles}. {\em\protect\JournalTitle{J. Fluid Mech.}}  \bvol{805},
  \pg{551--576}.

\bibitem[Grigoryev {\em et~al.\/}(2018)Grigoryev, B.V., Krivokorytov,
  Zhakhovsky, Dyachkov, Ilnitsky, Migdal, Inogamov, Vinokhodov, Kompanets,
  Sidelnikov, Krivtsun, Koshelev \& V.V.]{grigoryev_2018}
{\sc \au{Grigoryev, S.}, \au{B.V., L.}, \au{Krivokorytov, M.}, \au{Zhakhovsky,
  V.}, \au{Dyachkov, S.}, \au{Ilnitsky, D.}, \au{Migdal, K.}, \au{Inogamov,
  N.}, \au{Vinokhodov, A.}, \au{Kompanets, V.}, \au{Sidelnikov, Y.},
  \au{Krivtsun, V.}, \au{Koshelev, K.} \& \au{V.V., M.}} \yr{2018}
  \at{Expansion and fragmentation of a liquid-metal droplet by a short laser
  pulse}. {\em\protect\JournalTitle{Phys. Rev. Applied}}  \bvol{10}~(064009).

\bibitem[He {\em et~al.\/}(2017)He, Xin, Zhao, Chu, Xi, Shui, Lu \& Gu]{He_17}
{\sc \au{He, W.}, \au{Xin, J.}, \au{Zhao, Y.}, \au{Chu, G.}, \au{Xi, T.},
  \au{Shui, M.}, \au{Lu, F.} \& \au{Gu, Y.}} \yr{2017}  \at{Fragment size
  distribution statistics in dynamic fragmentation of laser shock-loaded tin}.
  {\em\protect\JournalTitle{AIP Advances}}  \bvol{7},  \pg{065306}.

\bibitem[Hecht(2002)]{hecht_optics_2002}
{\sc \au{Hecht, E.}} \yr{2002} {\em Optics\/}.  \publ{{Addison-Wesley}}.

\bibitem[Kafalas \& Ferdinand(1973)]{kafalas_fog_1973}
{\sc \au{Kafalas, P.} \& \au{Ferdinand, A.~P.}} \yr{1973}  \at{Fog {{Droplet
  Vaporization}} and {{Fragmentation}} by a 10.6-$\upmu$m {{Laser Pulse}}}.
  {\em\protect\JournalTitle{Applied Optics}}  \bvol{12}~(1),  \pg{29--33}.

\bibitem[Keller \& Kolodner(1954)]{keller_instability_1954}
{\sc \au{Keller, J.~B.} \& \au{Kolodner, I.}} \yr{1954}  \at{Instability of
  {{Liquid Surfaces}} and the {{Formation}} of {{Drops}}}.
  {\em\protect\JournalTitle{J. Appl. Phys.}}  \bvol{25}~(7),  \pg{918--921}.

\bibitem[Klein {\em et~al.\/}(2015)Klein, Bouwhuis, Visser, Lhuissier, Sun,
  Snoeijer, Villermaux, Lohse \& Gelderblom]{klein_drop_2015}
{\sc \au{Klein, A.~L.}, \au{Bouwhuis, W.}, \au{Visser, C.~W.}, \au{Lhuissier,
  H.}, \au{Sun, C.}, \au{Snoeijer, J.~H.}, \au{Villermaux, E.}, \au{Lohse, D.}
  \& \au{Gelderblom, H.}} \yr{2015}  \at{Drop {{Shaping}} by {{Laser}}-{{Pulse
  Impact}}}. {\em\protect\JournalTitle{Phys. Rev. Appl.}}  \bvol{3}~(4),
  \pg{044018}.

\bibitem[Klein {\em et~al.\/}(2017)Klein, Lohse, Versluis \&
  Gelderblom]{klein_apparatus_2017}
{\sc \au{Klein, A.~L.}, \au{Lohse, D.}, \au{Versluis, M.} \& \au{Gelderblom,
  H.}} \yr{2017}  \at{Apparatus to control and visualize the impact of a
  high-energy laser pulse on a liquid target}.
  {\em\protect\JournalTitle{Rev.Sci. Instrum.}}  \bvol{88},  \pg{095102}.

\bibitem[Kurilovich {\em et~al.\/}(2018)Kurilovich, De~Faria~Pinto, Torretti,
  Schupp, Scheers, Stodolna, Gelderblom, Eikema, Witte, Ubachs, Hoekstra \&
  O.O.]{kurilovich_2018}
{\sc \au{Kurilovich, D.}, \au{De~Faria~Pinto, T.}, \au{Torretti, F.},
  \au{Schupp, R.}, \au{Scheers, J.}, \au{Stodolna, A.}, \au{Gelderblom, H.},
  \au{Eikema, K.}, \au{Witte, S.}, \au{Ubachs, W.}, \au{Hoekstra, R.} \&
  \au{O.O., V.}} \yr{2018}  \at{Expansion dynamics after laser-induced
  cavitation in liquid tin microdroplets}. {\em\protect\JournalTitle{Phys. Rev.
  Applied}}  \bvol{10}~(054005).

\bibitem[Kurilovich {\em et~al.\/}(2016)Kurilovich, Klein, Torretti, Lassise,
  Hoekstra, Ubachs \& Versolato]{kurilovich_plasma_2016}
{\sc \au{Kurilovich, D.}, \au{Klein, A.~L.}, \au{Torretti, F.}, \au{Lassise,
  A.}, \au{Hoekstra, R.}, \au{Ubachs, W.and~Gelderblom, H.} \& \au{Versolato,
  O.~O.}} \yr{2016}  \at{Plasma {{Propulsion}} of a {{Metallic Microdroplet}}
  and its {{Deformation}} upon {{Laser Impact}}}.
  {\em\protect\JournalTitle{Phys. Rev. Appl.}}  \bvol{6}~(1),  \pg{014018}.

\bibitem[Lhuissier \& Villermaux(2011)]{lhuissier_2011}
{\sc \au{Lhuissier, H.} \& \au{Villermaux, E.}} \yr{2011}  \at{Destabilisation
  of an initially thick liquid sheet edge}. {\em\protect\JournalTitle{Phys.
  Fluids.}}  \bvol{23}~(091705).

\bibitem[Lhuissier \& Villermaux(2013)]{lhuissier_effervescent_2013}
{\sc \au{Lhuissier, H.} \& \au{Villermaux, E.}} \yr{2013}
  \at{`{{Effervescent}}' atomization in two dimensions}.
  {\em\protect\JournalTitle{Journal of Fluid Mechanics}}  \bvol{714},
  \pg{361--392}.

\bibitem[Lindinger {\em et~al.\/}(2004)Lindinger, Hagen, Socaciu, Bernhardt,
  W{\"o}ste, Duft \& Leisner]{lindinger_time-resolved_2004}
{\sc \au{Lindinger, A.}, \au{Hagen, J.}, \au{Socaciu, L.~D.}, \au{Bernhardt,
  T.~M.}, \au{W{\"o}ste, L.}, \au{Duft, D.} \& \au{Leisner, T.}} \yr{2004}
  \at{Time-resolved explosion dynamics of {{H}}\textsubscript{2}{{O}} droplets
  induced by femtosecond laser pulses}. {\em\protect\JournalTitle{Appl. Opt.}}
  \bvol{43}~(27),  \pg{5263--5269}.

\bibitem[Marpaung {\em et~al.\/}(2001)Marpaung, Kurniawan, Tjia \&
  Kagawa]{marpaung_comprehensive_2001}
{\sc \au{Marpaung, A.~M.}, \au{Kurniawan, H.}, \au{Tjia, M.~O.} \& \au{Kagawa,
  K.}} \yr{2001}  \at{Comprehensive study on the pressure dependence of shock
  wave plasma generation under {{TEA CO}}\textsubscript{2} laser bombardment on
  metal sample}. {\em\protect\JournalTitle{J. Phys. D: Appl. Phys.}}
  \bvol{34}~(5),  \pg{758--771}.

\bibitem[Masnavi {\em et~al.\/}(2011)Masnavi, Nakajima, Horioka, Araghy \&
  Endo]{masnavi_simulation_2011}
{\sc \au{Masnavi, M.}, \au{Nakajima, M.}, \au{Horioka, K.}, \au{Araghy, H.~P.}
  \& \au{Endo, A.}} \yr{2011}  \at{Simulation of particle velocity in a
  laser-produced tin plasma extreme ultraviolet source}.
  {\em\protect\JournalTitle{J. Appl. Phys.}}  \bvol{109}~(12),  \pg{123306}.

\bibitem[Ohl {\em et~al.\/}(2006)Ohl, Arora, Dijkink, Janve \&
  Lohse]{ohl_surface_2006}
{\sc \au{Ohl, C.-D.}, \au{Arora, M.}, \au{Dijkink, R.}, \au{Janve, V.} \&
  \au{Lohse, D.}} \yr{2006}  \at{Surface cleaning from laser-induced cavitation
  bubbles}. {\em\protect\JournalTitle{Appl. Phys. Lett.}}  \bvol{89}~(7),
  \pg{074102}.

\bibitem[Oron {\em et~al.\/}(1997)Oron, Davis \& Bankoff]{oron_long-scale_1997}
{\sc \au{Oron, A.}, \au{Davis, S.~H.} \& \au{Bankoff, S.~G.}} \yr{1997}
  \at{Long-scale evolution of thin liquid films}.
  {\em\protect\JournalTitle{Rev. Mod. Phys.}}  \bvol{69}~(3),  \pg{931--980}.

\bibitem[Peters {\em et~al.\/}(2013)Peters, Tagawa, Oudalov, Sun, Prosperetti,
  Lohse \& {van der Meer}]{peters_highly_2013}
{\sc \au{Peters, I.~R.}, \au{Tagawa, Y.}, \au{Oudalov, N.}, \au{Sun, C.},
  \au{Prosperetti, A.}, \au{Lohse, D.} \& \au{{van der Meer}, D.}} \yr{2013}
  \at{Highly focused supersonic microjets: Numerical simulations}.
  {\em\protect\JournalTitle{J. Fluid Mech.}}  \bvol{719},  \pg{587--605}.

\bibitem[Poulain {\em et~al.\/}(2018)Poulain, Villermaux \&
  Bourouiba]{poulain_ageing_2018}
{\sc \au{Poulain, S.}, \au{Villermaux, E.} \& \au{Bourouiba, L.}} \yr{2018}
  \at{Ageing and burst of surface bubbles}. {\em\protect\JournalTitle{J. Fluid
  Mech.}}  \bvol{851},  \pg{636--671}.

\bibitem[Reijers {\em et~al.\/}(2017)Reijers, Snoeijer \&
  Gelderblom]{reijers_pulses_2017}
{\sc \au{Reijers, S.}, \au{Snoeijer, J.} \& \au{Gelderblom, H.}} \yr{2017}
  \at{Droplet deformation by short laser-induced pressure pulses}.
  {\em\protect\JournalTitle{J. Fluid Mech.}}  \bvol{828},  \pg{374--394}.

\bibitem[Riboux \& Gordillo(2015)]{riboux_diameters_2015}
{\sc \au{Riboux, G.} \& \au{Gordillo, J.~M.}} \yr{2015}  \at{The diameters and
  velocities of the droplets ejected after splashing}.
  {\em\protect\JournalTitle{J. Fluid Mech.}}  \bvol{772},  \pg{630--648}.

\bibitem[Roisman {\em et~al.\/}(2006)Roisman, Horvat \&
  Tropea]{roisman_horvat_2006}
{\sc \au{Roisman, I.~V.}, \au{Horvat, K.} \& \au{Tropea, C.}} \yr{2006}
  \at{Spray impact: rim transverse instability initiating fingering and splash
  and description of a secondary spray}. {\em\protect\JournalTitle{Phys.
  Fluids}}  \bvol{18}~(102104).

\bibitem[Sigrist \& Kneub{\"u}hl(1978)]{sigrist_lasergenerated_1978}
{\sc \au{Sigrist, M.~W.} \& \au{Kneub{\"u}hl, F.~K.}} \yr{1978}
  \at{Laser-generated stress waves in liquids}. {\em\protect\JournalTitle{J.
  Acoust. Soc. Am.}}  \bvol{64}~(6),  \pg{1652--1663}.

\bibitem[Stan {\em et~al.\/}(2016)Stan, Milathianaki, Laksmono, Sierra,
  McQueen, Messerschmidt, Williams, Koglin, Lane, Hayes, Guillet, Liang,
  Aquila, Willmott, Robinson, Gumerlock, Botha, Nass, Schlichting, Shoeman,
  Stone \& Boutet]{stan_2016}
{\sc \au{Stan, C.~A.}, \au{Milathianaki, D.}, \au{Laksmono, H.}, \au{Sierra,
  R.~G.}, \au{McQueen, T.~A.}, \au{Messerschmidt, M.}, \au{Williams, G.~J.},
  \au{Koglin, J.~E.}, \au{Lane, T.~J.}, \au{Hayes, M.~J.}, \au{Guillet, S.
  A.~H.}, \au{Liang, M.}, \au{Aquila, A.~L.}, \au{Willmott, P.~R.},
  \au{Robinson, J.~S.}, \au{Gumerlock, K.~L.}, \au{Botha, S.}, \au{Nass, K.},
  \au{Schlichting, I.}, \au{Shoeman, R.~L.}, \au{Stone, H.~A.} \& \au{Boutet,
  S.}} \yr{2016}  \at{Liquid explosions induced by x-ray laser pulses}.
  {\em\protect\JournalTitle{Nat. Phys.}}  \bvol{12}~(10).

\bibitem[Tagawa {\em et~al.\/}(2012)Tagawa, Oudalov, Visser, Peters, {van der
  Meer}, Sun, Prosperetti \& Lohse]{tagawa_highly_2012}
{\sc \au{Tagawa, Y.}, \au{Oudalov, N.}, \au{Visser, C.~W.}, \au{Peters, I.~R.},
  \au{{van der Meer}, D.}, \au{Sun, C.}, \au{Prosperetti, A.} \& \au{Lohse,
  D.}} \yr{2012}  \at{Highly {{Focused Supersonic Microjets}}}.
  {\em\protect\JournalTitle{Phys. Rev. X}}  \bvol{2}~(3),  \pg{031002}.

\bibitem[Thoroddsen {\em et~al.\/}(2009)Thoroddsen, Takehara, Etoh \&
  Ohl]{thoroddsen_spray_2009}
{\sc \au{Thoroddsen, S.~T.}, \au{Takehara, K.}, \au{Etoh, T.~G.} \& \au{Ohl,
  C.-D.}} \yr{2009}  \at{Spray and microjets produced by focusing a laser pulse
  into a hemispherical drop}. {\em\protect\JournalTitle{Phys. Fluids}}
  \bvol{21}~(11),  \pg{112101}.

\bibitem[Utsunomiya {\em et~al.\/}(2010)Utsunomiya, Kajiwara, Nishiyama,
  Nagayama, Kubota \& Nakahara]{utsunomiya_laser_2010}
{\sc \au{Utsunomiya, Y.}, \au{Kajiwara, T.}, \au{Nishiyama, T.}, \au{Nagayama,
  K.}, \au{Kubota, S.} \& \au{Nakahara, M.}} \yr{2010}  \at{Laser ablation of
  liquid surface in air induced by laser irradiation through liquid medium}.
  {\em\protect\JournalTitle{Appl. Phys. A}}  \bvol{101}~(1),  \pg{137--141}.

\bibitem[Vernay {\em et~al.\/}(2015)Vernay, Ramos \& Ligoure]{vernay_free_2015}
{\sc \au{Vernay, C.}, \au{Ramos, L.} \& \au{Ligoure, C.}} \yr{2015}  \at{Free
  radially expanding liquid sheet in air: Time- and space-resolved measurement
  of the thickness field}. {\em\protect\JournalTitle{J. Fluid Mech.}}
  \bvol{764},  \pg{428--444}.

\bibitem[Villermaux(2007)]{villermaux_fragmentation_2007}
{\sc \au{Villermaux, E.}} \yr{2007}  \at{Fragmentation}.
  {\em\protect\JournalTitle{Ann. Rev. Fluid Mech.}}  \bvol{39}~(1),
  \pg{419--446}.

\bibitem[Villermaux \& Bossa(2011)]{villermaux_drop_2011}
{\sc \au{Villermaux, E.} \& \au{Bossa, B.}} \yr{2011}  \at{Drop fragmentation
  on impact}. {\em\protect\JournalTitle{J. Fluid Mech.}}  \bvol{668},
  \pg{412--435}.

\bibitem[Vledouts {\em et~al.\/}(2016)Vledouts, Quinard, Vandenberghe \&
  Villermaux]{vledouts_explosive_2016}
{\sc \au{Vledouts, A.}, \au{Quinard, J.}, \au{Vandenberghe, N.} \&
  \au{Villermaux, E.}} \yr{2016}  \at{Explosive fragmentation of liquid
  shells}. {\em\protect\JournalTitle{J. Fluid Mech.}}  \bvol{788},
  \pg{246--273}.

\bibitem[Vogel \& Venugopalan(2003)]{vogel_mechanisms_2003}
{\sc \au{Vogel, A.} \& \au{Venugopalan, V.}} \yr{2003}  \at{Mechanisms of
  {{Pulsed Laser Ablation}} of {{Biological Tissues}}}.
  {\em\protect\JournalTitle{Chem. Re.}}  \bvol{103}~(2),  \pg{577--644}.

\bibitem[Wang \& Xu(2001)]{wang_thermoelastic_2001}
{\sc \au{Wang, X.} \& \au{Xu, X.}} \yr{2001}  \at{Thermoelastic wave induced by
  pulsed laser heating:}. {\em\protect\JournalTitle{Appl. Phys. A Mater. Sci.
  Proces.}}  \bvol{73}~(1),  \pg{107--114}.

\bibitem[Wang {\em et~al.\/}(2018)Wang, Dandekar, Bustos, Poulain \&
  Bourouiba]{bourouiba_2018}
{\sc \au{Wang, Y.}, \au{Dandekar, R.}, \au{Bustos, N.}, \au{Poulain, S.} \&
  \au{Bourouiba, L.}} \yr{2018}  \at{Universal rim thickness in unsteady sheet
  fragmentation}. {\em\protect\JournalTitle{Phys. Rev. Lett.}}
  \bvol{120}~(204503).

\bibitem[Xu {\em et~al.\/}(2007)Xu, Barcos \& Nagel]{xu_splashing_2007}
{\sc \au{Xu, L.}, \au{Barcos, L.} \& \au{Nagel, S.~R.}} \yr{2007}
  \at{Splashing of liquids: {{Interplay}} of surface roughness with surrounding
  gas}. {\em\protect\JournalTitle{Phys. Rev. E}}  \bvol{76}~(6),  \pg{066311}.

\end{thebibliography}

\end{document}